\documentclass[pre,twocolumn,showpacs,showkeys,amsmath,amssymb,floatfix,eqsecnum]{revtex4}
\usepackage{graphicx}
\usepackage{bm}

\begin{document}

\title{Thermodynamic Casimir effects involving interacting field
  theories with zero modes}

\author{Daniel Gr{\"u}neberg} \author{H.~W.  Diehl}
\affiliation{%
  Fachbereich Physik, Universit{\"a}t Duisburg-Essen, D-47048 Duisburg,
  Germany}

\date{\today}

\begin{abstract}
  Systems with an $O(n)$ symmetrical Hamiltonian are considered in a
  $d$-dimensional slab geometry of macroscopic lateral extension and
  finite thickness $L$ that undergo a continuous bulk phase transition
  in the limit $L\to\infty$. The effective forces induced by thermal
  fluctuations at and above the bulk critical temperature
  $T_{c,\infty}$ (thermodynamic Casimir effect) are investigated below
  the upper critical dimension $d^*=4$ by means of field-theoretic
  renormalization group methods for the case of periodic and
  special-special boundary conditions, where the latter correspond to
  the critical enhancement of the surface interactions on both
  boundary planes. As shown previously [\textit{Europhys.\ Lett.}
  \textbf{75}, 241 (2006)], the zero modes that are present in Landau
  theory at $T_{c,\infty}$ make conventional RG-improved perturbation
  theory in $4-\epsilon$ dimensions ill-defined. The revised expansion
  introduced there is utilized to compute the scaling functions of the
  excess free energy and the Casimir force for temperatures $T\geq
  T_{c,\infty}$ as functions of $\mathsf{L}\equiv L/\xi_\infty$, where
  $\xi_\infty$ is the bulk correlation length. Scaling functions of
  the $L$-dependent residual free energy per area are obtained whose
  $\mathsf{L}\to0$ limits are in conformity with previous results for
  the Casimir amplitudes $\Delta_C$ to $O(\epsilon^{3/2})$ and display
  a more reasonable small-$\mathsf{L}$ behavior inasmuch as they
  approach the critical value $\Delta_C$ monotonically as
  $\mathsf{L}\to 0$.  Extrapolations to $d=3$ for the Ising case $n=1$
  with periodic boundary conditions are in fair agreement with Monte
  Carlo results. In the case of special-special boundary conditions,
  extrapolations to $d=3$ are hampered by the fact that the one-loop
  result for the inverse finite-size susceptibility becomes negative
  for some values of $\mathsf{L}$ when $\epsilon\gtrsim 0.83$.
\end{abstract}


\maketitle

\section{Introduction}
\label{sec:intro}

When a classical or quantum fluid, or an $n$-vector magnet with
$n=1$, $2$, $3$ is confined by macroscopic bodies such as two
parallel plates, walls, surfaces, or interfaces, of area $A$, its free
energy $F$ depends upon the distance $L$ between these boundary planes.
The $L$ dependence implies a force
\begin{equation}
   \label{eq:FCdef}
{\mathcal F}_C(T,L)=-k_BT\,\frac{\partial
  f_{\text{ex}}(T,L)}{\partial L}
\end{equation}
between the plates, where $f_{\text{ex}}=(F/k_BT)-Lf_\text{b}$ is the reduced
excess free energy per unit area, $f_\text{b}$ is the reduced bulk free
energy density, and the limit $A\to\infty$ has been
taken \cite{FdG78,rem:casitdrev,Kre94,Kre99,BDT00}. By analogy with the
familiar Casimir force \cite{Cas48} produced by vacuum fluctuations of
the electromagnetic field between two metallic plates (and slight
abuse of language), ${\mathcal F}_C$ is conventionally called
``thermodynamic Casimir force.''

Provided long-range interactions are either absent or negligible, this
force decays exponentially for separations $L\gtrsim \xi_\infty(T)$, where $\xi_\infty(T)$ is
the bulk correlation length. Near a continuous phase transition in $d$
bulk dimensions, $\xi_\infty(T)$ diverges as $|T-T_{c,\infty}|^{-\nu}$
at the bulk critical temperature $T_{c,\infty}$. Therefore, the
Casimir force ${\mathcal F}_C(T,L)$ extends to distances $L$ much
larger than the microscopic scale $a$ ($\simeq$ radius of atoms,
lattice constant).

Writing
\begin{equation}
  \label{eq:delfex}
  f_{\text{ex}}(T,L)=f_s(T)+ f_{\text{res}}(T,L)\;,
\end{equation}
let us decompose the excess free energy $f_{\text{ex}}$ into an
$L$~independent surface part $f_s(T)\equiv f_{\text{ex}}(T,\infty)$
and a residual finite-size contribution $f_{\text{res}}(T,L)$. The
latter behaves as
  \begin{equation}
  \label{eq:cadef} 
f_{\text{res}}(T_{c,\infty},L) \mathop{\approx}\limits_{L\to
  \infty}\Delta_C\,L^{-(d-1)} 
\end{equation}
at the bulk critical point $T=T_{c,\infty}$, and hence produces the
long-ranged effective force
\begin{equation}
\label{eq:FCTc}
  \frac{\mathcal{F}_{C}(T_{c,\infty},L)}{k_BT_{c,\infty}}
  \mathop{\approx}\limits_{L\to \infty} (d-1)\,\Delta_C\,L^{-d} \;.
\end{equation}
Here $\Delta_C$, the so-called Casimir amplitude \cite{rem:casitdrev}, is a
universal quantity, which depends on the bulk universality class of
the phase transition considered and gross properties of the boundary
plates, but is independent of microscopic details.

Such thermodynamic Casimir forces have been the subject of much
interest recently \cite{KD91,KD92a,KD92b,GC99,GSGC06}. Clear experimental
evidence for their existence has been found in the thinning of wetting
layers of liquid $^4$He as a function of temperature on approaching 
the lambda line \cite{GC99,GSGC06}.

Near the bulk critical point the residual free energy density and
Casimir force are expected to have the scaling forms
\begin{equation}
  \label{eq:scffintro}
  f_{\text{res}}(T,L) \approx L^{-(d-1)}\,\Theta(L/\xi_\infty)
\end{equation}
and
\begin{equation}\label{eq:scfFCintro}
  \frac{\mathcal{F}_{C}(T_{c,\infty},L)}{k_BT}\approx
  L^{-d}\,\Xi(L/\xi_\infty)\;,
\end{equation}
where $\Theta$ and
\begin{equation}
  \label{eq:ThetaXi}
  \Xi(\mathsf{L})=(d-1)\,\Theta(\mathsf{L})
  -\mathsf{L}\,\Theta^\prime(\mathsf{L})  
\end{equation}
should be universal functions of $\mathsf{L}\equiv L/\xi_\infty$.
These expectations rest on the assumption that $\xi_\infty$ and $L$
are large compared to other lengths, which means, in particular, that
the symmetry breaking field $h$ vanishes and long-range interactions
are either absent or negligible.

In studies of the Casimir effect in QED, matter usually is taken into
account only through the choice of appropriate boundary conditions on
the surfaces of macroscopic bodies. Hence they involve \emph{free}
field theories under given boundary conditions. Systematic theoretical
investigations of the Casimir effect at \emph{critical points} are a
much greater challenge in that one has to deal with \emph{interacting}
field theories in finite and bounded systems \cite{Die86a,Die97}.

A first fairly detailed study of the thermodynamic Casimir effect was
made about 15 years ago by Krech and Dietrich (KD) for the $\phi^4$
theory on a slab $\mathbb{R}^{d-1}\times [0,L]$ of thickness $L$
\cite{KD91,KD92a}. Building on Symanzik's work \cite{Sym81} in the 80s
and the simultaneously emerging field-theory approach to critical
behavior of systems with boundaries
\cite{DD80,DD81a,DD81b,DD83a,Die86a,Die97}, these authors considered
five different boundary conditions $\wp$, namely, periodic
($\wp=\text{per}$), antiperiodic ($\wp=\mathrm{ap}$), and the three
nonequivalent combinations $(D,D)$, $(D,\mathrm{sp})$, and
$(\mathrm{sp},\mathrm{sp})$ of Dirichlet ($D$) and special ($\mathrm{sp}$)
boundary conditions on the slab's two boundary planes. Here the former
(D) means $\bm{\phi}=\bm{0}$ as usual, while the latter
($\mathrm{sp}$) is the case of a Robin boundary condition
$\partial_n\bm{\phi}=\mathring{c}\,\bm{\phi}$ for which $\mathring{c}$
takes the special value $\mathring{c}_{\mathrm{sp}}$ corresponding to
the critical enhancement of the surface interactions on the respective
boundary plane.

Restricting themselves to temperatures $T\geq T_{c,\infty}$, KD
performed two-loop calculations for $4-\epsilon$ dimensional slabs
under these boundary conditions $\wp$ and determined the $\epsilon$
expansions of the Casimir amplitudes $\Delta_C^{(\wp)}$ as well as
those of the corresponding scaling functions $\Theta^{(\wp)}$ to first
order in $\epsilon$.

In a recent paper with Shpot \cite{DGS06}, we have shown that
conventional renormalization-group (RG) improved perturbation theory,
on which both Symanzik's \cite{Sym81} and KD's \cite{KD91,KD92a}
analyses are based, becomes ill-defined at $T_{c,\infty}$ beyond
two-loop order due to infrared singularities for those boundary
conditions that involve a zero mode at $T_{c,\infty}$ in Landau
theory.  This applies to both $\wp=\mathrm{per}$ and
$\wp=(\mathrm{sp},\mathrm{sp})$. To remedy these deficiencies, we
performed a reorganization of field theory such that the resulting
RG-improved perturbation theory remained meaningful at $T_{c,\infty}$.
It was found that the small-$\epsilon$ expansions of the corresponding
Casimir amplitudes $\Delta^{(\wp)}_C$ involve fractional powers
$\epsilon^{k/2}$, with $k\geq 3$, and powers of $\ln\epsilon$.
Furthermore, explicit results for these series to order
$\epsilon^{3/2}$ were given.

In this paper we will utilize this approach to compute the scaling
functions $\Theta^{(\wp)}$, and hence
$\Xi^{(\wp)}$, for $\wp=\mathrm{per}$ and $\mathrm{(sp,sp)}$ to the same order of RG-improved perturbation
theory.  The results are consistent with, and reproduce those of
\cite{DGS06} when $T=T_{c,\infty}$.

\begin{figure}[t]
  \centering
\includegraphics[width=0.85\columnwidth,clip]{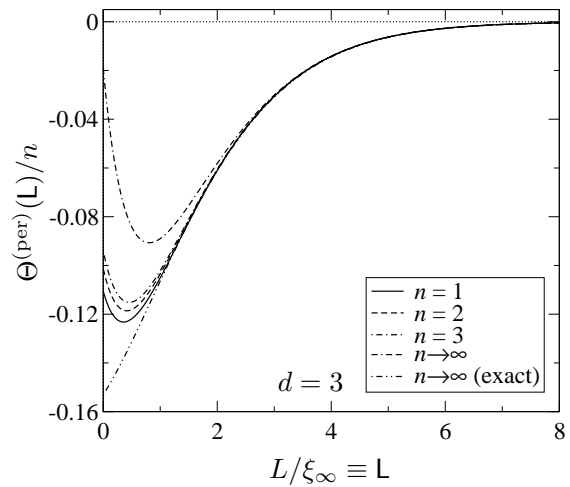}
\caption{Scaling functions $\Theta^{(\mathrm{per})}$ for
  $n=1,2,3,\infty$ obtained by extrapolating the $O(\epsilon)$ results
  of \cite{KD92a} to $d=3$, compared with the exact large-$n$ result
  for $d=3$ \cite{Dan96,BDT00}.} \label{fig:Thetaextr}
\end{figure}

Let us note that KD's two-loop results for these boundary conditions,
though well-defined down to $T_{c,\infty}$, gave clear indications of
existing problems. To see this, consider the scaling functions
$\Theta^{(\mathrm{per})}$ for $n=1,2,3,\infty$ displayed in
Fig.~\ref{fig:Thetaextr}, which were obtained by extrapolating their
$O(\epsilon)$ results to $d=3$. 

The behavior of these curves at small $\mathsf{L}$ differs in a
qualitative fashion from that of the exact scaling function for
$n=\infty$ and $d=3$, which follows from the exact solution of the
mean spherical model under periodic boundary conditions
\cite{Dan98,BDT00,DDG06}. Unlike the latter, which decreases
monotonically to its critical value
\begin{equation}
  \label{eq:DeltaSMperd3}
\Delta_C^{(\mathrm{per,SM})}=\frac{-2\,\zeta(3)}{5\pi} \simeq  -0.15305
\end{equation}
at $\mathsf{L}=0$, the former go through a minimum at small
$\mathsf{L}>0$ and then increase as $\mathsf{L}\to 0$.  Such a minimum
is neither expected at $\mathsf{L}>0$ nor in conformity with the Monte
Carlo results of Ref.~\cite{DK04} and announced more recent ones
\cite{Huc07,Huc:tbp,VGMD07}. Note also that as $n$ increases,
the extrapolations actually move away from the exact $n=\infty$
curve since the deviations at small $\mathsf{L}$ get bigger. 

A second problem was pointed to by KD: Since for finite $L$ no phase
transition takes place at $T_{c,\infty}$, the free energy per unit
area must be an analytic function of temperature at $T_{c,\infty}$,
which imposes conditions on the small-$\mathsf{L}$ behavior of the
scaling functions $\Theta^{(\wp)}(\mathsf{L})$ (which will be recalled
in Sec.~\ref{sec:genprop}). KD found their $O(\epsilon)$ results to be
consistent with these conditions only in the considered cases
of non-zero-mode boundary conditions $\wp=\mathrm{ap}$, $(D,D)$, and
$(D,\mathrm{sp})$.  In the remaining cases of the zero-mode boundary
conditions $\wp=\mathrm{per}$ and $(\mathrm{sp},\mathrm{sp)}$, these
conditions turned out to be violated by terms of first order in
$\epsilon$.

The results our approach yields for the scaling functions
$\Theta^{(\mathrm{per})}$ do better in two regards. First, the
small-$\mathsf{L}$ behavior is improved inasmuch as the Casimir
amplitudes $\Delta_C^{(\mathrm{per})}$ are approached in a
monotonically decreasing manner as $\mathsf{L}\to0$. Second, the order
of the terms violating the analyticity condition is increased from
$O(\epsilon)$ to $O(\epsilon^{3/2})$. In the case of sp-sp boundary
conditions, our results raise questions whose answers might require a
generalization of our analysis in which the surface enhancement
variables are allowed to vary. As we shall see, the one-loop
expression for the scaling function of the inverse finite-size
susceptibility becomes negative in a small interval of
$\mathsf{L}=L/\xi_\infty$ when evaluated at $\epsilon=1$. This
probably simply means that this extrapolation to $d=3$ is not
sufficiently accurate. In any case, this violation of a necessary
stability condition of the disordered phase is a problem even for KD's
original $O(\epsilon)$ results.

The remainder of this paper is organized as follows. In the next
section we specify the model utilized in our analysis --- the $\phi^4$
theory in slab geometry. We briefly recapitulate the general
fluctuating Robin boundary conditions it involves, its
renormalization, the fixed points that are relevant for the subsequent
analysis, and the renormalization of its free energy. In
Sec.~\ref{sec:revftapp} we first recall the conventional theory of the
Casimir effect based on RG improved perturbation theory in
$4-\epsilon$ bulk dimensions, and then discuss the problems into which
it runs when a zero mode appears in Landau theory. This is followed by
a detailed exposition of how these problems can be overcome through an
appropriate reformulation of field theory. In Sec.~\ref{sec:calcfe}
our results for the residual free energies and their
scaling functions are presented. In Sec.~\ref{sec:compsmres} we employ
the solution of the mean spherical model under periodic boundary
conditions \cite{Dan96,BDT00,DDG06} for $d<4$ to show that our
small-$\epsilon$ results are in conformity with these exact ones in
the limit $n\to \infty$. A brief summary and discussion of our work is
given in Sec.~\ref{sec:concl}. Finally, there are four appendixes in
which technical details are described.

\section{Continuum model, boundary conditions, and background}
\label{sec:fp}

\subsection{Definition of model}

We consider a $d$-dimensional slab of finite thickness $L$ occupying the
region $\mathfrak{V}=\mathbb{R}^{d-1}\times [0,L]$ of $d$-dimensional
space. Let $x_j,\;j=1,\dotsc,d,$ be Cartesian coordinates, with
$x_d\equiv z$ taken along the finite direction. We write the position
vector $\bm{x}=(x_1,\dotsc,x_d)$ as $\bm{x}=(\bm{y},z)$, where
$\bm{y}=(x_1,\dotsc,x_{d-1})$ is the component along the slab. 

The Hamiltonians of the $\phi^4$ models we are concerned with are sums
of a bulk and a boundary term,
\begin{equation}
  \label{eq:Hform}
\mathcal{H}[\bm{\phi}]=
\int_{\mathfrak{V}}\mathcal{L}_{\mathfrak{V}}(\bm{x})\,dV 
+\int_{\partial\mathfrak{V}}
\mathcal{L}_{\partial\mathfrak{V}}(\bm{x})\,dA \;,
\end{equation}
where $\mathcal{L}_{\mathfrak{V}}(\bm{x})$ and
$\mathcal{L}_{\partial\mathfrak{V}}(\bm{x})$ depend on 
$\bm{\phi}(\bm{x})$ and its derivatives.

We either consider periodic or free boundary conditions along the
$z$~direction. In the first case, where
\begin{equation}
  \label{eq:papbc}
  \bm{\phi}(\bm{x}+L\,\hat{\bm{e}}_d)=  \bm{\phi}(\bm{x})\;,
\end{equation}
there is no boundary, $\partial\mathfrak{V}=\emptyset$, and the
boundary term $\int_{\partial\mathfrak{V}}\ldots$ is absent. In the
case of free boundary conditions, the boundary
$\partial\mathfrak{V}=\mathfrak{B}_1\cup \mathfrak{B}_2$ is the union
of $\mathfrak{B}_1$, the $z=0$~plane, and $\mathfrak{B}_2$, the
$z=L$~plane.

The bulk density is always given by
\begin{equation}
  \label{eq:Lb}
  \mathcal{L}_{\mathfrak{V}}[\bm{\phi}]= \frac{1}{2}\,
  \sum_{\alpha=1}^n(\nabla\phi_\alpha)^2 
  +\frac{\mathring{\tau}}{2}\,\phi^2 +\frac{\mathring{u}}{4!}\,\phi^4\;,
\end{equation}
where $\bm{\phi}(\bm{x})=(\phi_\alpha(\bm{x}))$ is the
$n$-component order parameter field and $\phi$ denotes its
absolute value $|\bm{\phi}|$.

The boundary density we utilize when considering free boundary conditions
reads
\begin{equation}
  \label{eq:L1}
  \mathcal{L}_{\partial\mathfrak{V}}[\bm{\phi}]=
  \frac{\mathring{c}(\bm{x})}{2}\,\phi(\bm{x})^2\;,
\end{equation}
where  $\mathring{c}(\bm{x})$, the surface enhancement variable, is allowed to
have different values on $\mathfrak{B}_1$ and $\mathfrak{B}_2$, i.e.,
\begin{equation}
  \label{eq:c12}
  \mathring{c}(\bm{x})=
  \begin{cases}
    \mathring{c}_1& \text{for }\bm{x}\in\mathfrak{B}_1,\\
\mathring{c}_2& \text{for }\bm{x}\in\mathfrak{B}_2.
  \end{cases}
\end{equation}

\subsection{Boundary conditions}

Using well-known arguments \cite{Die86a,Die97}, one concludes from the
boundary terms in the classical equations of motion
$\delta\mathcal{H}=0$ that the derivative
$\partial_n\bm{\phi}$ along the inner normal $\bm{n}$ on
$\partial\mathfrak{V}$ satisfies
\begin{equation}
  \label{eq:bc}
  \partial_n\phi_\alpha(\bm{x})=\mathring{c}_j\,\phi_\alpha(\bm{x})
  \quad\text{for }\bm{x}\in 
  \mathfrak{B}_j\;.
\end{equation}
This is a boundary condition for Landau theory, which holds beyond it in
an operator sense (inside of averages).

\subsection{Renormalization of correlation functions}

To absorb the ultraviolet (uv) singularities of the ($N+M$)-point
cumulant functions 
\begin{equation}\label{eq:GNMdef}
G_{\alpha_1,\dotsc,\beta_M}^{(N,M)}(\bm{x}_1,\dotsc,\bm{y}_M)
=\bigg\langle\prod_{j=1}^N\phi_{\alpha_j}(\bm{x}_j)
\prod_{k=1}^M\phi_{\beta_k}(\bm{y}_k)\bigg\rangle^{\mathrm{cum}}
\end{equation}
involving $N$ interior points $\bm{x}_j\notin\partial\mathfrak{V}$ and
$M$ boundary points $\bm{y}_k\in\partial\mathfrak{V}$ for dimensions
$d\leq 4$, bulk and boundary counterterms are needed, which can be
chosen to correspond to the reparametrizations
\begin{eqnarray} \label{eq:bulkreps}
  \bm{\phi}&=&Z_\phi^{1/2}\,\bm{\phi}_R\;,\nonumber\\
\mathring{\tau}-
  \mathring{\tau}_{c,\infty}\equiv\delta\mathring{\tau}&=&Z_\tau\,\mu^2\tau\;,\nonumber\\ 
\mathring{u}\,N_d&=&\mu^\epsilon\, Z_u\,u\;, 
  \end{eqnarray}
and
\begin{eqnarray}\label{eq:surfreps}
    \delta\mathring{c}_j\equiv\mathring{c}_j-\mathring{c}_{\mathrm{sp}}&=&\mu Z_c\,c_j\;,\nonumber\\
    \bm{\phi}|_{\partial\mathfrak{B}}&=&(Z_\phi Z_1)^{1/2}\,
    \bm{\phi}|_{\partial\mathfrak{B}}^R\;.
\end{eqnarray}
Here $\bm{\phi}|_{\partial\mathfrak{B}}$ means $\bm{\phi}(\bm{y}_k)$
at a boundary point $\bm{y}_k$, and $\mu$ is an arbitrary momentum
scale.  Further, $\delta\mathring{\tau}$ is the deviation of $\mathring{\tau}$ from
$\mathring{\tau}_{c,\infty}$, the critical-point value of $\mathring{\tau}$ of the bulk
system. In a theory regularized by a large-momentum cutoff $\Lambda$,
$\mathring{\tau}_{c,\infty}$ would diverge $\sim \Lambda^2$. We prefer to use
dimensional regularization; then $\mathring{\tau}_{c,\infty}$ vanishes in
perturbation theory.  The renormalization factors $Z_\phi$, $Z_\tau$,
and $Z_u$ are standard bulk quantities. The renormalization factors
$Z_1$ and $Z_c$ are properties of the semi-infinite system that
results  in the limit $L\to\infty$ when $c_1=c_2$ with $|c_1|<\infty$.

We choose the factor that is absorbed in the renormalized coupling
constant as \cite{rem:corrlength}
\begin{eqnarray}
  \label{eq:Nd}
  N_d&=&\frac{2\,\Gamma(3-d/2)}{(d-2)(4\pi)^{d/2}}\nonumber \\ &=&
\frac{1}{16\pi^2}\left[1+\frac{1-C_E+\ln(4\pi)}{2}\,\epsilon
  +O(\epsilon^2)\right]\;, \qquad
\end{eqnarray} 
where $C_E$ is the Euler-Mascheroni constant. It differs from the one advocated by Schloms and Dohm (see, e.g.,
\cite{SD89} and its references) by a trivial factor of $2$. 
Ours agrees to zeroth order in $\epsilon$ with
($2^{-d}\pi^{-d/2}$), the one employed in Ref.~\cite{Die86a} and by KD.
Therefore, all of the above bulk and surface renormalization factors
$Z_\phi$,\ldots,$Z_1$ remain the same as in \cite{Die86a,KD92a} when
determined by minimal subtraction of poles at $\epsilon=0$.  Explicit
two-loop expressions for these functions can be found in
Eqs.~(3.42a--c) and Eqs.~(3.66a,b) of \cite{Die86a}, or in
Refs.~\cite{DD81b,DD83a}. The advantage of our choice of $N_d$ is to
simplify the resulting expressions for renormalized one-loop bulk
vertex functions while leaving the renormalization factors of
\cite{Die86a} and KD unchanged. 

The quantity $\mathring{c}_{\mathrm{sp}}$ is the special value of
$\mathring{c}_1$ corresponding to the critical enhancement of the
surface interactions in a semi-infinite system with surface plane
$\mathfrak{B}_1$ --- i.e., the value at which the so-called special
transition occurs (provided the surface dimension $d-1$ is large
enough to allow long-range surface order above $T_{c,\infty}$).
Analogously to $\mathring{\tau}_{c,\infty}$, it would diverge
$\sim\Lambda$ as $\Lambda\to\infty$ in a cut-off regularized theory,
but vanishes in a perturbative approach based on dimensional
regularization \cite{DS94,DS98,rem:nonanalytcsp}.

\subsection{Fixed points}\label{sec:fpts}

Let $u^*$ be the infrared-stable zero of the beta function
$\beta_u(u)\equiv\mu\partial_\mu|_0u$, where $\partial_{\mu}|_0$ means a
derivative at fixed bare parameters $\mathring{u}$, $\mathring{\tau}$, $\mathring{c}_1$, and $\mathring{c}_2$
of the theory. In the enlarged space $\{\tau,u,c_1,c_2\}$ of bulk and
surface variables, the RG yields fixed points  on the hyperplane
$(\tau,u)=(0,u^*)$ located at the 9 pairs $(c_1^*,c_2^*)$ of the
fixed-point values
\begin{equation}
    \label{eq:fpvalues}
    c_j^*=
    \begin{cases}
      c^*_{\mathrm{ord}}&=\infty\;,\\
c^*_{\mathrm{sp}}&=0\;,\\
c^*_{\mathrm{ex}}&=-\infty\;.
    \end{cases}
\end{equation}
These values pertain to the fixed points describing the critical
behavior at the ordinary, special, and extraordinary transitions of
the semi-infinite system. Each one of these fixed points is specified
by a pair $(\kappa_1,\kappa_2)$ with $\kappa_1,\,\kappa_2=\mathrm{ord}$,
$\mathrm{sp}$, $\mathrm{ex}$ of the respective surface universality
classes. Universal finite-size quantities such as the Casimir
amplitudes $\Delta_C^{(\wp)}$ and the scaling functions
$\Theta^{(\wp)}$, $\Xi^{(\wp)}$ generally are different, depending on
the basin of attraction of the fixed point $(\kappa_1,\kappa_2)$ to
which they belong. Recall that for $\kappa_j=\mathrm{ord}$ the cumulants
(\ref{eq:GNMdef}) satisfy the Dirichlet boundary condition
$\lim_{\bm{x}_k\to \mathfrak{B}_j}G^{(N,0)}=0$. Thus the universality classes
$(\mathrm{ord},\mathrm{ord})$, $(\mathrm{ord},\kappa)$, and
$(\kappa,\mathrm{ord})$ with $\kappa\neq \mathrm{ord}$ can equivalently be
labeled as $(D,D)$, $(D,\kappa)$, and $(\kappa,D)$, respectively. We
shall continue to employ this convention.

\subsection{Renormalization of free energy}\label{sub:renf}

The counterterms implied by the reparametrizations~(\ref{eq:bulkreps})
and (\ref{eq:surfreps}) are sufficient to absorb the uv singularities
of the cumulants~(\ref{eq:GNMdef}). However, the free energy requires
additional additive counterterms \cite{Die86a,KD92a}. They can be
chosen to be independent of $L$ \cite{Sym81,Die86a}. We therefore add
to the Hamiltonian defined by Eqs.~(\ref{eq:Hform})--(\ref{eq:c12})
a contribution
\begin{equation}
  \mathcal{A}_{\mathrm{add}}=\int_{\mathfrak{V}}C_{\mathfrak{V}}(\mathring{\tau},\mathring{u})\,dV
  +\sum_{j=1}^2\int_{\mathfrak{B}_j}
  C_{\partial\mathfrak{V}}(\mathring{\tau},\mathring{u},\mathring{c}_j)\,dA \;,
\end{equation}
where $C_{\mathfrak{V}}$ is a polynomial in $\mathring{\tau}$ of degree 2,
$C_{\partial\mathfrak{V}}$ is a polynomial of degree one in $\mathring{\tau}$ and
degree 3 in $\mathring{c}_j$, whose coefficients depend on $\mathring{u}$, but neither
on $L$ nor on the position $\bm{x}$. The coefficients (power series in
$\mathring{u}$) are fixed as follows \cite{rem:KD}. Let $ T^{\leq
  4}_{\mathrm{NP}}f(\mathring{\tau})$ denote 
the Taylor series expansions of the function $f$ to \emph{second}
order in $\mathring{\tau}$ (4th order in $\mathring{\tau}^{1/2}$), and
\begin{eqnarray}
T^{\leq 3}_{\mathrm{NP}}g(\mathring{\tau},\mathring{c})&=&\sum_{\substack{j,k\\ 0
\leq 2j+k\leq 3}}\frac{1}{j!\,k!}\,\frac{\partial^{j+k}g}{\partial
   \mathring{\tau}^j\,\partial\mathring{c}^k}\Big|_{\mathrm{NP}}\, \nonumber\\
&&\strut\times 
 (\mathring{\tau}-\mathring{\tau}_{\mathrm{NP}})^j\,(\mathring{c}-\mathring{c}_{\mathrm{NP}})^{k}\;.\quad
\end{eqnarray}
be the corresponding expansion of $g$ in $\mathring{\tau}$ and $\mathring{c}$  to
orders $j$ and $k$ with $2j+k\leq 3$ about the
normalization point $(\mathring{\tau},\mathring{c})=(\mathring{\tau}_{\mathrm{NP}},\mathring{c}_{\mathrm{NP}})$
\cite{rem:sfdd}. We choose
\begin{eqnarray}
  \mathring{\tau}_{\mathrm{NP}}\equiv
  \mathring{\tau}|_{\tau=1}&=&\mathring{\tau}_{c,\infty}+\mu^2\,Z_\tau\;,\nonumber\\ 
\mathring{c}_{\mathrm{NP}}\equiv
  \mathring{c}|_{c=1}&=&\mathring{c}_{\mathrm{sp}}+\mu\,Z_c\;,
\end{eqnarray}
and define the dimensionless renormalized bulk free energy density
$f_{b,R}$ by 
\begin{equation}\label{eq:fbsubtr}
  \mu^{-d}\,f_{b,R}(\tau,u)=f_{b}(\mathring{\tau},\mathring{u})-T^{\leq 4}_{\mathrm{NP}}f_{b}(\mathring{\tau},\mathring{u})\;.
\end{equation}

The excess surface free energy density $f_s(\mathring{\tau},\mathring{u},\mathring{c}_1,\mathring{c}_2)$ of
the infinitely thick film is a sum of contributions
$f_s(\mathring{\tau},\mathring{u},\mathring{c}_1)$ associated with the respective semi-infinite
systems bounded on one side by $\mathfrak{B}_j$; i.e.,
\begin{eqnarray}
  f_s(\mathring{\tau},\mathring{u},\mathring{c}_1,\mathring{c}_2)=f_s(\mathring{\tau},\mathring{u},\mathring{c}_1)+f_s(\mathring{\tau},\mathring{u},\mathring{c}_2)\;.
\end{eqnarray}
We define the dimensionless renormalized analogs of the latter by
\begin{equation}
   \mu^{-(d-1)}\,f_{s,R}(\tau,u,c)=f_s(\mathring{\tau},\mathring{u},\mathring{c})
   -T^{\leq 3}_{\mathrm{NP}}f_s(\mathring{\tau},\mathring{u},\mathring{c})\;.
\end{equation}

By construction, these renormalized bulk and surface free energy
densities satisfy the normalization conditions
\begin{equation}
  f_{b,R}\big|_{\mathrm{NP}}\equiv f_{b,R}(1,u)=0 =\frac{\partial
    f_{b,R}}{\partial \tau}\Big|_{\mathrm{NP}} =\frac{\partial^2
    f_{b,R}}{\partial \tau^2}\Big|_{\mathrm{NP}} 
\end{equation}
and
\begin{equation}
  \frac{\partial^{j+k}f_{s,R}}{\partial\tau^j\,\partial
    c^k}\Big|_{\mathrm{NP}}\equiv
  \frac{\partial^{j+k}f_{s,R}}{\partial\tau^j\,\partial 
    c^k}(1,u,1)=0 \;,\;\;0\leq 2j+k\leq 3\;, 
\end{equation}
respectively. The renormalization functions $C_{\mathfrak{V}}$ and
$C_{\partial\mathfrak{V}}$ are fixed by these requirements.

The renormalized excess surface free energy density
$f_{s,R}(\tau,u,c_1,c_2)$ one obtains from the action
$\mathcal{H}+\mathcal{A}_{\mathrm{add}}$ upon insertion of the
reparametrizations is uv finite. Since $C_{\mathfrak{V}}$ and
$C_{\partial\mathfrak{V}}$ are independent of $L$, the subtractions
they provide cancel in the residual free energy
$f_{\mathrm{res}}(L;\mathring{\tau},\mathring{u},\mathring{c}_1,\mathring{c}_2)$ of the film. Accordingly, its
dimensionless renormalized counterpart
\begin{equation}
  f_{\mathrm{res},R}(\mu L;\tau,u,c_1,c_2)=\mu^{-(d-1)}f_{\mathrm{res}}(L;\mathring{\tau},\mathring{u},\mathring{c}_1,\mathring{c}_2) 
\end{equation}
satisfies a \emph{homogeneous} RG equation, whereas both $f_{b,R}$ and
$f_{s,R}$ satisfy inhomogeneous ones. 

Following the notation conventions of Ref.~\cite{Die86a}, we introduce
the beta function $\beta_u=\mu\partial_\mu|_0u$, the RG functions
$\eta_\kappa=\mu\partial_\mu|_0\ln Z_\kappa$,
$\kappa=\phi,\tau,u,c,1$, and the operator
\begin{equation}\label{eq:Dmu}
  \mathcal{D}_\mu=\mu\partial_\mu+\beta_u\partial_u
  -(2+\eta_\tau)\partial_\tau -(1+\eta_c)\sum_{j=1}^2
  c_j\partial_{c_j} \;.
\end{equation}
Then the RG equation of $f_{\mathrm{res},R}$ can be written as
\begin{equation}\label{eq:RGfres}
  [\mathcal{D}_\mu+(d-1)]f_{\mathrm{res},R}(\mu L;\tau,u,c_1,c_1)=0\;.
\end{equation}
Note that the RG functions $\beta_u$ and $\eta_\kappa$ are either bulk
quantities (such as $\beta_u(\epsilon,u)$, $\eta_\tau(u)$) or
properties of semi-infinite systems such as $\eta_c(u)$. In accordance
with Ref.~\cite{Die86a}, we have chosen them independent of $c_j$
and $\tau$ (fixing them by minimal subtraction of poles in
$\epsilon$). Explicit two-loop expressions for these functions can be
found in Eqs.~(3.75a)--(3.76b) of this reference.

The RG equation~(\ref{eq:RGfres}) can be solved in a standard fashion
by means of characteristics. Upon setting $\mu=1$ and choosing the
scale parameter $\ell$ of the transformation $\mu\to\mu\ell$ equal to
$1/\xi_\infty$, the inverse bulk correlation length, we see that the
residual free energy density, on sufficiently long length scales,
takes the finite-size scaling form
\begin{equation}
  f_{\mathrm{res},R}(L;\tau,u,c_1,c_1)\approx
  L^{-(d-1)}\,\Theta(L/\xi_\infty;c_1\,\xi_\infty^\Phi,c_2\,
  \xi_\infty^\Phi)\;.  
\end{equation}
Here $\Phi$ is the surface crossover exponent of the special
transition. The scaling function $\Theta$ is universal up to the
nonuniversal amplitude of $\xi_\infty$ and the nonuniversal metric
factor associated with $c_1$ and $c_2$ \cite{rem:Ec}; it is
given by 
\begin{equation}\label{eq:Thetafresrel}
  \Theta(\mathsf{L};\mathsf{c}_1,\mathsf{c}_2)
  =\mathsf{L}^{d-1}\,
  f_{\mathrm{res},R}(\mathsf{L};1,u^*,\mathsf{c}_1,\mathsf{c}_2) \;.  
\end{equation}

From it the functions $\Theta^{(\wp)}(\mathsf{L})$ with
$\wp=(\mathrm{sp},\mathrm{sp})$, $(D,D)$, and $(\mathrm{sp},D)$ follow by
setting $(\mathsf{c}_1,\mathsf{c}_2)$ to the respective fixed-point
values $(0,0)$, $(\infty,\infty)$, and $(0,\infty)$. For example,
\begin{equation}
  \Theta^{(\mathrm{sp},\mathrm{sp})}(\mathsf{L})
  =\mathsf{L}^{d-1}\,f_{\mathrm{res},R}(\mathsf{L};1,u^*,0,0)\;.
\end{equation}

For reasons explained in the introduction, we shall mainly be
concerned with the cases of periodic and $(\mathrm{sp},\mathrm{sp})$ boundary
conditions. 

\section{Revised field theory approach}\label{sec:revftapp}

\subsection{Infrared problems due to zero modes}

We now turn to the problem of computing the scaling functions
$\Theta^{(\wp)}$ and $\Xi^{(\wp)}$ for $\wp=\mathrm{per},
(\mathrm{sp},\mathrm{sp})$ by means of RG-improved perturbation theory.
Only the case $T\geq T_{c,\infty}$ will be considered.

The free propagator can be written as
\begin{equation}
  \label{eq:fprop}
   G_{L}^{(\wp)}(\bm{x};\bm{x}'|\mathring{\tau})=\int_{\bm{p}}^{(d-1)}
  \sum_m\frac{\langle z|m\rangle\langle m|
    z'\rangle}{p^2+k_m^2+\mathring{\tau}}\,e^{i\bm{p}\cdot(\bm{y}-\bm{y}')}\;,
\end{equation}
where
\begin{equation}
  \int_{\bm{p}}^{(d-1)}\equiv\int \frac{d^{d-1}p}{(2\pi)^{d-1}}
\end{equation}
is a convenient shorthand for a normalized $d-1$~dimensional momentum
integral. Further, $|m\rangle=|m\rangle^{(\wp)}$ are eigenstates given
by
\begin{equation}
  \label{eq:per}
  \langle z|m\rangle^{(\mathrm{per})}=\frac{\exp(ik_mz)}{\sqrt{L}}\;,
  \quad k_m=\frac{2\pi m}{L}\;,\;m\in\mathbb{Z}\;,
\end{equation}
and
\begin{eqnarray}
  \label{eq:sp}
  \langle z|m\rangle^{(\mathrm{sp},\mathrm{sp})}&=&
  \frac{1}{\sqrt{L}}\begin{cases}1
    &\text{for }m=0\;,\\[\smallskipamount]
\sqrt{2}\,\cos(k_mz)&\text{for }m\in\mathbb{N}\;,
  \end{cases} \nonumber\\[\medskipamount]
  k_m&=&m\pi/L\,,\;\;m=0,1,\dotsc,\infty\;,
\end{eqnarray}
respectively. For either boundary condition the mode with $m=0$ and
$\bm{p}=\bm{0}$ becomes massless at $\tau=0$. 

\begin{figure}[htb]
  \centering
 \includegraphics[scale=0.8,clip]{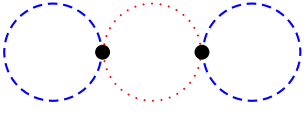}
  \caption{(color online) Infrared divergent contribution to the free energy. The
    dashed full blue lines represent the $k_m\neq 0$~part of the free
    propagator~(\ref{eq:fprop}); the dotted
    red lines denote its $k_m= 0$~part. The blue subgraphs approach a
    finite $L$-dependent limit as $\tau\to 0$; the red dashed
    subgraphs varies as a negative power of $\tau$ and hence is
    infrared singular \cite{DGS06}.}\label{fig:irdivgraph}
\end{figure}

In their calculation of $\Delta^{(\wp)}_C$ directly at $T_{c,\infty}$,
KD therefore subtracted the contribution from the $m=0$ mode to avoid
infrared problems, the rationale being that the subtracted one-loop
contribution is formally independent of $L$ so that it does not
contribute to the Casimir force. Computing the one- and two-loop
graphs at $\tau\geq 0$ \cite{KD92a,DGS06}, one finds that the
contributions from the $k_0=0$ modes vary as a positive power of $\tau$
and hence vanish as $\tau\to 0$. However, at the three-loop level this
is no longer the case because one encounters infrared divergent
contributions of the form depicted in Fig.~\ref{fig:irdivgraph}. 
Thus conventional RG-improved perturbation theory is ill-defined at
$T_{c,\infty}$.

The origin of this problem is that Landau theory yields sharp
transitions for both the bulk and the film system at the same critical
value $\mathring{\tau}=0$. It is thus of a similar kind as encountered in the
study of finite-size effects of systems that are finite in all, or in
all but one, direction under periodic boundary conditions
\cite{BZJ85,RGJ85}. As discussed in Ref.~\cite{DGS06}, the remedy is
to separate the $k_0=0$ mode and construct an effective field theory for
the $k_0=0$ part of the order parameter.

\subsection{Construction of effective zero-mode action}

To this end we write
\begin{equation}\label{eq:phidec}
  \bm{\phi}(\bm{x})=\sum_{m}\bm{\phi}_m(\bm{y})\,\langle z|m\rangle=
L^{-1/2}\,\bm{\varphi}(\bm{y})+\bm{\psi}(\bm{y},z)\;,
\end{equation}
decomposing the order parameter into its component along
$\bm{\phi}_0(\bm{y})=\bm{\varphi}(\bm{y})$ and a remaining
$k_m\neq 0$ contribution $\bm{\psi}(\bm{y},z)$ with
\begin{equation}
  \int_0^Ldz\,\bm{\psi}(\bm{y},z)=0\;.
\end{equation}
Tracing out $\bm{\psi}$ defines us a ($d-1$)-dimensional effective field
theory with the Hamiltonian 
\begin{eqnarray}\label{eq:heffdef}
  \mathcal{H}_{\mathrm{eff}}[\bm{\varphi}]&=&-\ln\mathrm{Tr}_{\bm{\psi}}
  e^{-\mathcal{H}[L^{-1/2}\bm{\varphi}+\bm{\psi}]}\nonumber\\
&=&\frac{F_{\bm{\psi}}}{k_BT}+\mathcal{H}[L^{-1/2}\bm{\varphi}]-\ln\langle
  e^{-\mathcal{H}_{\mathrm{int}}[\bm{\varphi},\bm{\psi}]}
  \rangle_{\bm{\psi}}\;.\qquad 
\end{eqnarray}
Here $F_{\bm{\psi}}$, defined by
\begin{equation}
  \exp(-F_{\bm{\psi}}/k_BT)=\mathrm{Tr}_{\bm{\psi}}
  \exp\left(-\mathcal{H}\big[\bm{\psi}\big]\right),
\end{equation}
is the free energy due to the $k_m\neq 0$ modes. Further,
\begin{equation}
  \label{eq:Hint}
  \mathcal{H}_{\mathrm{int}}[\bm{\varphi},\bm{\psi}]\equiv
  \int_{\mathfrak{V}} dV\, 
  \Big[\frac{\mathring{u}}{4\,L}\,\varphi^2\,\psi^2
    +\frac{\mathring{u}}{6\sqrt{L}}\,(\bm{\varphi}\cdot\bm{\psi})\,\psi^2\Big]
\end{equation}
is the interaction part, and
\begin{equation}
  \label{eq:heff0}
  \mathcal{H}\big[L^{-1/2}\bm{\varphi}\big]
  =\int dA\,
  \bigg[\frac{1}{2}\sum_{j=1}^{d-1}\Big(\frac{\partial \bm{\varphi}}{\partial x_j}\Big)^2
    +\frac{\mathring{\tau}}{2}\,\varphi^2 +\frac{\mathring{u}
    }{4!\,L}\,\varphi^4\bigg]\;. 
\end{equation}

Computing the last term on the right-hand side of
Eq.~(\ref{eq:heffdef}) in a loop expansion gives
\begin{equation}\label{eq:heff}
   \mathcal{H}_{\mathrm{eff}}[\bm{\varphi}]=\frac{F_{\bm{\psi}}}{k_BT}
   +\mathcal{H}[L^{-1/2}\bm{\varphi}]
   +\mathcal{H}^{[1]}_{\mathrm{eff}}[\bm{\varphi}] +\ldots
\end{equation}
with
\begin{eqnarray}
  \label{eq:heff1}
\mathcal{H}_{\mathrm{eff}}^{[1]}[\bm{\varphi}]&=&\frac{1}{2}\,\mathrm{Tr}
 \ln\Big[\openone+\frac{\mathring{u}}{6L}\,G^{(\wp)}_{L,\psi}
   \big(\delta_{\alpha\beta}\,\varphi^2
     +2\varphi_\alpha\varphi_\beta\big)\Big]\nonumber\\
&=& -
\,\,\raisebox{-4pt}{\includegraphics[scale=1,clip]{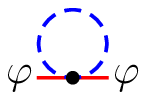}}\,\,
-\,\,\raisebox{-10pt}{\includegraphics[scale=1,clip]{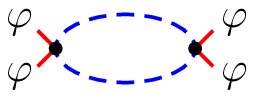}}
\,\,+O(\mathring{u}^3)\;,\nonumber\\
\end{eqnarray}
where the dashed blue lines (color online) represent free $\psi$-propagators
\begin{equation}\label{eq:psiprop}
  G^{(\wp)}_{L,\psi}(\bm{x};\bm{x}'|\mathring{\tau})=\int_{\bm{p}}^{(d-1)}
  \sum_{m\neq 0}\frac{\langle z|m\rangle\langle m|
    z'\rangle}{p^2+k_m^2+\mathring{\tau}}\,e^{i\bm{p}\cdot(\bm{y}-\bm{y}')}
\end{equation}
and the red bars indicate $\varphi$~legs.

Writing
\begin{eqnarray}\label{eq:gvarphi}
  \lefteqn{\delta_{\alpha\beta}\,g_\varphi^{-1}(\bm{y}-\bm{y}')= \frac{\delta^2\mathcal{H}_{\mathrm{eff}}[\bm{\varphi}]
  }{\delta\varphi_\alpha(\bm{y})\,\delta\varphi_\beta(\bm{y}^\prime)}
  \bigg|_{\bm{\varphi}=\bm{0}} }&&\nonumber\\ &=&
  \delta_{\alpha\beta}\left[(-\partial^2
    +\mathring{\tau})\delta(\bm{y}-\bm{y}')-\sigma_\varphi(\bm{y}-\bm{y}')\right]
 \;,
\end{eqnarray}
where $\partial^2$ means the Laplacian $\sum_{j=1}^{d-1}\partial_j^2$
in $\mathbb{R}^{d-1}$,
and
\begin{equation}
  \gamma^{(k)}_{\alpha_1,\dotsc,\alpha_k}(\bm{y}_1,\dotsc,\bm{y}_k)=
  \frac{\delta^k\mathcal{H}_{\mathrm{eff}}[\bm{\varphi}] 
  }{\delta\varphi_{\alpha_1}(\bm{y}_1)\dotsm\delta
    \varphi_{\alpha_k}(\bm{y}_k)} 
  \bigg|_{\bm{\varphi}=\bm{0}} \;,
\end{equation} 
we introduce the propagator $g_\varphi$ associated with the
$\varphi^2$ term of $\mathcal{H}_{\mathrm{eff}}[\bm{\varphi}]$ and the
corresponding self-energy $\sigma_\varphi$ as well as the vertices
$\gamma^{(k)}$ of the effective action
$\mathcal{H}_{\mathrm{eff}}[\bm{\varphi}]$. Though not indicated here, all
these quantities depend on $L$ and the boundary condition $\wp$.

\begin{figure}[htb]
  \centering
  \includegraphics[scale=0.8,clip]{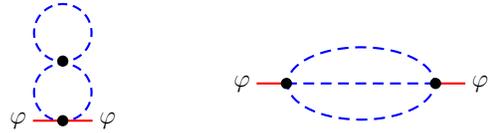}
  \caption{(color online) Two-loop contributions to $\sigma_\varphi$. The left graph is
    local, the one on the right-hand side is nonlocal.}\label{fig:twoloopsigma}
\end{figure}

The first graph in the second line of Eq.~(\ref{eq:heff1}) is the
one-loop contribution to $\sigma_\varphi$. It is local in
$\bm{y}$-space. As can be seen from Fig.~\ref{fig:twoloopsigma}, both
local and nonlocal contributions appear beyond one-loop order.
The second graph in the lower line of Eq.~(\ref{eq:heff1}) is the
nonlocal one-loop contribution to $\gamma^{(4)}$. Evidently, vertices
$\gamma^{(k)}$ of arbitrary even order $k$ are generated through the
coupling to the $k_m\neq 0$ modes. 

\subsection{RG-improved perturbation theory}\label{sec:RGipt}

Now suppose the bulk critical point is approached so that $\xi_\infty$
becomes large. Then the vertices $\gamma^{(k)}$ cannot be computed by
perturbation theory below the upper critical dimension $d^*=4$.
However, for arbitrary small $\tau>0$ we can employ the RG to map to a
system with a minimal length scale on the order of $\xi_\infty$, and
then employ perturbation theory. The vertex functions $\gamma^{(k)}$
are expected to decay as a function of the relative differences
$y_{ij}=|\bm{y}_i-\bm{y}_j|$ on the scale of $\xi_\infty$.

The renormalized counterparts  $\gamma_R^{(k)}$ of these vertices
satisfy the RG equations \cite{rem:RGvert}
\begin{equation}
 \left( \mathcal{D}_\mu-\frac{N}{2}\,\eta_\phi\right)\gamma_R^{(k)}=0\;.
\end{equation}
Solving them in a standard fashion, one finds that the Fourier
transforms
$\hat{\gamma}^{(k)}\,(2\pi)^{d-1}\,\delta(\sum_{j=1}^k\bm{p}_j)$ of
these functions on sufficiently large length scales take the scaling forms 
\begin{equation}
  \hat{\gamma}_R^{(2k)}(\{\bm{p}_i\})\approx
  \mu^{k\eta}\,L^{1-d+(d-3+\eta)k}\,
  X_{2k}^{(\wp)}(\{\bm{p}_i\xi_\infty\};L/\xi_\infty)\;,
\end{equation}
where $\eta$ is a standard bulk critical exponent, while $\xi_\infty$
is the second-moment bulk correlation length. The latter is defined in
the conventional manner in terms of the bulk vertex function
$\tilde{\Gamma}^{(2)}_\mathrm{b}(\bm{q})=1/\tilde{G}_\mathrm{b}^{(2)}(\bm{q})$
of the $\phi^4$ theory in $d$-dimensional momentum space or its
position-space back transform $G^{(2)}_\mathrm{b}(\bm{x})$ via
\begin{equation}
  \label{eq:xiinfdef}
\xi_\infty^{2} \equiv \left.\left[\frac{\partial}{\partial
      q^{2}} \ln\tilde{\Gamma}^{(2)}_\mathrm{b}(\bm{q})\right]
\right|_{\bm{q}=\bm{0}}=
\frac{1}{2d}\, \frac{\int d^dx\,x^2\,G_\mathrm{b}^{(2)}(\bm{x})}{\int
  d^dx\,G^{(2)}_\mathrm{b}(\bm{x})} \;.
\end{equation}

Let us verify explicitly to first order in $u^*=O(\epsilon)$ that
RG-improved perturbation theory yields such scaling behavior.
Consider, for example, $\gamma^{(2)}_R=g_{\varphi,R}^{-1}$. The first
graph in the second line of Eq.~(\ref{eq:heff1}) is the $O(\mathring{u})$
contribution to $\hat{\sigma}_\varphi(p)$. Introducing
\begin{equation}\label{eq:Iwpj}
  I^{(\wp)}_j(L;\mathring{\tau}) \equiv \int_{0}^{L}\frac{dz}{L}
  \Big[G_{L,\psi}^{(\wp)}(\bm{x};\bm{x}|
  \mathring{\tau})\Big]^{j}\;,\;\; j=1,2\;,
\end{equation}
we have
\begin{equation}\label{eq:sigphi}
\hat{\sigma}_\varphi(p)=-\mathring{u}
\frac{n+2}{6}\,I^{(\wp)}_1(L;\mathring{\tau}) +O(\mathring{u}^{2}) \;.
\end{equation}
The integrals $I^{(\wp)}_j(L;\mathring{\tau})$ are computed in
Appendix~\ref{app:reqintegr}. The results for $I^{(\mathrm{per})}_1$ and
$I^{(\mathrm{sp},\mathrm{sp})}_1$ are
\begin{equation}\label{eq:I1per}
I^{(\mathrm{per})}_1(L;\mathring{\tau})=
\frac{A_{d-1}}{L}\,\mathring{\tau}^{(d-3)/2}-A_{d}\,
\mathring{\tau}^{(d-2)/2} 
 + \frac{2\,Q_{d,2}(\mathring{\tau}L^{2})}{\mathring{\tau}L^{d}}
\end{equation}
and
\begin{equation}\label{eq:I1sp}
I^{\mathrm{(sp,sp)}}_1(L;\mathring{\tau})
=I^{(\mathrm{per})}_1(2L;\mathring{\tau})\;,
\end{equation} 
where 
\begin{equation}\label{eq:Ad}
A_d=\frac{2\,N_d}{4-d}=-(4\pi)^{-d/2}\,\Gamma(1-d/2)\;.
\end{equation}
Here
$Q_{d,2}(r)$ is a special one of the functions defined by
\begin{equation}\label{eq:definitionQdsigma}
Q_{d,\sigma}(r)\equiv\frac{r}{2} \left[\sum_{k\in2\pi\mathbb{Z}}
  \int_{\bm{p}}^{(d-1)}
  -\int_{\bm{q}=(\bm{p},k)}^{(d)}\right]\frac{q^{\sigma-2}}{q^{2}+r}\;,
\end{equation}
where $\bm{p}$ and $k$ are the ($d-1$)-parallel and one-dimensional
perpendicular components of the wave vector $\bm{q}=(\bm{p},k)$.
The properties of these functions are analyzed and discussed in
Appendix~\ref{app:Qd2}, where we compute them for the required
parameter values of $d$ and $\sigma$. Plots of the functions
$Q_{d,2}(r)$ with $d=4$ and 6 are displayed in Fig.~\ref{fig:Qd2}
(Appendix~\ref{app:Qd2}). 

To facilitate subsequent comparisons with KD's results, let us note how the
$Q_{d,2}(r)$ are related to the functions
\begin{equation}\label{eq:ga}
  g_{a,b}(z)\equiv \frac{1}{a} \int_1^\infty dt\,
\frac{(t^2-1)^a\,\ln^b(t^2-1)}{e^{2zt}-1}
\end{equation}
utilized by these authors. As shown in Appendix~\ref{app:Qdsigma},
one has
\begin{equation}
  \label{eq:Qd2ga}
    Q_{d,2}(r)=\frac{2^{1-d}\,\pi^{(1-d)/2}\,
    r^{d/2}}{\Gamma[(d-3)/2)]}\, g_{\frac{d-3}{2},0}(\sqrt{r}/2)\;.
\end{equation}

Using the results~(\ref{eq:I1per}) and (\ref{eq:I1sp}),
and expressing $g_{\varphi,R}^{-1}=Z_\phi\,g_\varphi^{-1}$ in terms of
the renormalized variables $\tau$ and $u$, one finds that the pole
$\sim \epsilon^{-1}$ cancels. The resulting renormalized expression is
easily evaluated at the fixed-point value $u=u^*$. It  conforms
with the scaling form
\begin{equation}\label{eq:scfgvarphi}
[\hat{g}_{\varphi,R}^{(\wp)}(p,\tau,L)]^{-1}\approx
\mu^{\eta}\,L^{\eta-2}\,X_2^{(\wp)}(p\,\xi_\infty,L/\xi_\infty)\;. 
\end{equation}
and yields for the scaling functions the $\epsilon$~expansions
\begin{eqnarray}
X_2^{(\mathrm{per})}(\mathsf{p},\mathsf{L})
&=&(\mathsf{p}^2+1)\mathsf{L}^2  +\frac{n+2}{n+8}\,\epsilon
\,\big[2\pi\, \mathsf{L}\nonumber\\ &&\strut 
  +16\pi^2\, Q_{4,2}(\mathsf{L}^2)/\mathsf{L}^2\big] +O(\epsilon^2)\quad
\end{eqnarray}
and
\begin{eqnarray}
    X_2^{\mathrm{(sp,sp)}}(\mathsf{p},\mathsf{L})&=&(\mathsf{p}^2+1)
  \mathsf{L}^2  +\frac{n+2}{n+8}\,
  \epsilon\,\big[\pi\,\mathsf{L}\nonumber\\ &&\strut 
  +\pi^2\,Q_{4,2}(4\mathsf{L}^2)/\mathsf{L}^2\big]
  +O(\epsilon^2)\;.\qquad 
\end{eqnarray}

A few comments are in order here. 

(i) The above results imply that
$[g^{(\wp)}_{\varphi,R}(0,\tau,L)]^{-1}$ does not vanish at $\tau=0$
when $L<\infty$. Using the fact that $\eta=O(\epsilon^2)$ and the
small-$r$ behavior of $Q_{4,2}(r)$ implied by Eq.~(\ref{eq:Qd2smally})
yields
\begin{eqnarray}
  \label{eq:shiftper}
[g^{(\mathrm{sp},\mathrm{sp})}_{\varphi,R}(0,0,L)^*]^{-1}&=&
 [4\,g^{(\mathrm{per})}_{\varphi,R}(0,0,L)^*]^{-1}+O(\epsilon^2)
  \nonumber\\ &=&
\epsilon\,\frac{n+2}{n+8}\,
\frac{\zeta(2)}{L^{2}}+O(\epsilon^2)\;,\qquad 
\end{eqnarray}
where the asterisk indicates evaluation at $u=u^*$.

The physical meaning of this result is obvious. The coupling of the
${k_0=0}$~mode to the $k_m\neq 0$ modes has produced  an $L$-dependent shift
of the temperature at which $\varphi$ becomes critical, making
$\varphi$ noncritical at $T_{c,\infty}$ when $L<\infty$.

(ii) Verifying the scaling form (\ref{eq:scfgvarphi}) to higher orders
in $\epsilon$ and the appearance of a nontrivial exponent $\eta$ by
extending RG-improved perturbation theory to $O[(u^*)^2]$ or higher is
in principle straightforward.

(iii) It is instructive to see what our procedure yields for boundary
conditions such as $\wp=(D,D)$, $(D,\mathrm{sp})$, and $(\mathrm{ap})$
where Landau theory does not involve a zero-mode at $T_{c,\infty}$. In
those cases we have $\bm{\varphi}\equiv \bm{0}$ and
$\bm{\phi}=\bm{\psi}$. Accordingly, Eq.~(\ref{eq:heff}) simply yields
$F/k_BT\equiv F_{\psi}/k_BT$ for the reduced free energy. It is therefore
clear that for those non-zero-mode boundary conditions conventional
expansions in integer powers of $\epsilon$ will result for the Casimir
force, the scaling functions $\Xi^{(\wp)}$ and $\Theta^{(\wp)}$, and
similar quantities for $T\geq T_{c,\infty}$, which must be in
accordance with KD's results to $O(\epsilon)$.

(iv) The procedure utilized above of constructing the action of an
effective lower-dimensional field theory by integrating out modes via
RG-improved perturbation theory that do not become critical for
$T=T_{c,\infty}$ at zero-loop order is similar to the one employed in
the study of static and dynamic finite-size effects in systems that
are finite in all, or in all but one, directions
\cite{BZJ85,Die87,Gol87,NZ-J87,RD96}. In the latter cases one arrives
for small deviations $\epsilon=d^*-d>0$ from the upper critical
dimension $d^*=4$ at expansions in powers of $\epsilon^{1/2}$ and
$\epsilon^{1/3}$, respectively. The main difference between these
cases and ours is that a \emph{sharp transition to a low-temperature
  phase with long-range order} is ruled out for the former because
they involve  systems of finite extent along $d$ or $d-1$ Cartesian axes (and the
presumed short-range interactions). By contrast, in the case of the
slab geometry considered here, such a sharp transition should occur
for finite thickness $L$ at a shifted temperature
$T_{c,L}<T_{c,\infty}$ whenever $d-1$, the effective dimensionality,
is sufficiently large for such a long-range ordered low-temperature
phase to occur. (Evidently $d-1$ must exceed $d_*(n)$, the lower
critical dimension, which is $d^*(1)=1$ in the Ising case $n=1$, and
$d_*(n>1)=2$, depending on whether a discrete $\mathbb{Z}_2$ or
continuous $O(n)$ symmetry gets spontaneously broken.) When no sharp
transition is possible, one expects a rounded one at a shifted
pseudo-critical temperature (see, e.g., Refs.~\cite{Fis71,Bar83}). The
case of $d=3$ and $n=2$, corresponding to an $XY$-model on a slab or
liquid $^4$He film below the bulk $\lambda$-line $T_\lambda$, is
exceptional in that a transition of Kosterlitz-Thouless type to a
low-temperature phase with quasi-long-range order is expected to occur
for finite $L$.

(v) That the coupling of the $k_0=0$~mode $\bm{\varphi}$ to the
${k\neq 0}$~modes $\bm{\psi}$ produces an $L$-dependent mass gap for
$g_\varphi^{-1}$ is crucial for making RG-improved perturbation theory
well-defined at $T_{c,\infty}$. However, it must be emphasized that
such a perturbative approach using $u^*=O(\epsilon)$ as expansion
parameter by itself \emph{must not be expected} to give a proper
description of the ($d-1$~dimensional) critical behavior at $T_{c,L}$!
One way to see this is to note that the bare $\varphi^4$ coupling
constant appearing in $\mathcal{H}_{\mathrm{eff}}[\bm{\varphi}]$ is
$\mathring{u}/L$.  To make it dimensionless we must multiply by the
$(5-d)$th power of a length. An appropriate one is $\xi_L$, the finite-size
analog of $\xi_\infty$, defined by
\begin{eqnarray}
  \label{eq:xiL}
  \xi_L^{2} &\equiv& \left.\left[\frac{\partial}{\partial
      p^{2}} \ln\hat{\Gamma}^{(2)}_{\varphi\varphi}(\bm{p})\right]
\right|_{\bm{p}=\bm{0}}\nonumber \\ &=&
\frac{1}{2(d-1)}\, \frac{\int d^{d-1}y\,y^2\,\langle\bm{\varphi}(\bm{y})\cdot\bm{\varphi}(\bm{0})\rangle^{\text{cum}}}{\int
  d^{d-1}y\,\langle\bm{\varphi}(\bm{y})\cdot\bm{\varphi}(\bm{0})\rangle^{\text{cum}}} \;,\qquad
\end{eqnarray}
where $\hat{\Gamma}_{\varphi\varphi}(\bm{p})$ denotes the full
$\varphi\varphi$ vertex function in the space of ($d-1$)-dimensional
momenta $\bm{p}$.

The appropriate dimensionless coupling
constant therefore is $\xi_L^{5-d}\mathring{u}/L$, which diverges as
$\xi_L\to\infty$ whenever $d<5$.  In accordance with general
expectations we thus see that the appropriate smallness parameter for
analyzing the $d-1$~dimensional critical behavior at $T_{c,L}$ by
means of a dimensionality expansion is $5-d$ rather than $\epsilon$.
Constructing a RG approach that is reliable both at $T_{c,\infty}$ and
$T_{c,L}$ and capable of describing the crossover from $d$ to
$d-1$~dimensional critical behavior is a nontrivial problem, which has
so far not been solved in a satisfactory fashion and is beyond the
scope of this paper.

\section{Calculation of free energies and scaling functions}\label{sec:calcfe} 
According to Eq.~(\ref{eq:heff}), the reduced bare free energy density per
unit area
\begin{equation}
  \label{eq:fLdef}
  f_L^{(\wp)}=\lim_{A\to\infty}\frac{F}{A k_BT}
\end{equation}
is a sum 
\begin{equation}\label{eq:fLfpsiFvarphi}
  f_L^{(\wp)}(\mathring{\tau})=f_\psi^{(\wp)}(L;\mathring{\tau})+f_\varphi^{(\wp)}(L;\mathring{\tau})
\end{equation}
of a contribution $f_\psi^{(\wp)}(L;\mathring{\tau})$ from the $k_m\neq 0$ modes
and a remainder, which we denote as $f_\varphi^{(\wp)}$. We first
consider the non-zero mode contribution $f_\psi^{(\wp)}$.

\subsection{Non-zero mode contribution to the free energy}\label{sec:fpsi}

A standard loop expansion yields
\begin{equation}
  f_\psi^{(\wp)}(L;\mathring{\tau})= f_{\psi,[1]}^{(\wp)}(L;\mathring{\tau})+
  f_{\psi,[2]}^{(\wp)}(L;\mathring{\tau})+O(3\text{-loops}) 
\end{equation}
with
\begin{eqnarray}
f_{\psi,[1]}^{(\wp)}(L;\mathring{\tau})&=&\frac{n}{2}  \sum_{k_m\neq 0}
\int_{\bm{p}}^{(d-1)}\ln(p^2+k_m^{2}+\mathring{\tau})\nonumber\\
&=&f_{\psi,[1]}^{(\wp)}(L;0) + 
L\,\frac{n}{2}\,J^{(\wp)}(L;\mathring{\tau})
\label{eq:fpsi1}\end{eqnarray}
and
\begin{equation}
  f_{\psi,[2]}^{(\wp)}(L;\mathring{\tau})=\mathring{u} L\frac{n(n+2)}{4!}\,I_2^{(\wp)}(L;\mathring{\tau})\;,
\label{eq:fpsi2}
\end{equation}
where
\begin{equation}\label{eq:Jwp}
  J^{(\wp)}(L;\mathring{\tau})=\int_0^{\mathring{\tau}} I_1^{(\wp)}(L;t)\,dt\;.
\end{equation}
The $\mathring{\tau}=0$~contributions $f_{\psi,[1]}^{(\wp)}(L;0)$ are computed in
Appendix~\ref{app:f10}. The results are in accordance with those of
KD. Expressed in terms of the familiar one-loop
values 
\begin{equation}\label{eq:oneloopCasi}
  \Delta_{C,[1]}^{(\mathrm{per})}=2^d\,\Delta_{C,[1]}^{(\mathrm{sp},\mathrm{sp})}
  =-n\pi^{-d/2}\,\Gamma(d/2)\,\zeta(d)
\end{equation}
of the Casimir amplitudes, they can be written as
\begin{equation}
  \label{eq:fpsizeroP}
  f_{\psi,[1]}^{(\wp)}(L;0)= f_{\psi,0}^{(\wp)}
+L^{-(d-1)}\Delta_{C,[1]}^{(\wp)}\;.
\end{equation}
Here $f_{\psi,0}^{(\wp)}$ are the cut-off and $L$ dependent quantities
defined by Eqs.~(\ref{eq:fpsi0wpapp})--(\ref{eq:fswppsi0});
they vanish in dimensional regularization.

The integrals $I^{(\wp)}_2$ and $J^{(\wp)}$ are worked out in
Appendix~\ref{app:reqintegr}. The results are given in
Eqs.~(\ref{eq:I2perres}), (\ref{eq:I2spspres}), (\ref{eq:Jperres}),
and (\ref{eq:Jspspres}). Inserting them into Eqs.~(\ref{eq:fpsi1}) and
(\ref{eq:fpsi2}) gives
\begin{eqnarray}\label{eq:fpsi1perres}
f_{\psi,[1]}^{(\mathrm{per})}(L;\mathring{\tau}) & = &
f_{\psi,0}^{(\mathrm{per})}+n\bigg[\frac{A_{d-1}}{d-1}\,
\mathring{\tau}^{(d-1)/2}-L\frac{A_{d}}{d}\,\mathring{\tau}^{d/2}\nonumber
\\ 
 &  & -\frac{4\pi
   Q_{d+2,2}(\mathring{\tau}L^{2})}{\mathring{\tau}L^{d+1}}\bigg],
\end{eqnarray}
\begin{eqnarray}\label{eq:fpsi2perres}
f_{\psi,[2]}^{(\mathrm{per})}(L;\mathring{\tau}) & = &
\mathring{u}L\frac{n(n+2)}{4!}
\bigg[\frac{A_{d-1}}{L}\,\mathring{\tau}^{(d-3)/2}\nonumber \\ 
 &  & -A_{d}\,\mathring{\tau}^{(d-2)/2}
 +\frac{2Q_{d,2}(\mathring{\tau}L^{2})}{\mathring{\tau}L^{d}}\bigg]^{2},\quad
\end{eqnarray}
and
\begin{eqnarray}\label{eq:fpsi1spres}
f_{\psi,[1]}^{(\mathrm{sp},\mathrm{sp})}(L;\mathring{\tau}) & = &
f_{\psi,0}^{(\mathrm{sp},\mathrm{sp})} +n\bigg[\frac{1}{2}\frac{A_{d-1}}{d-1}\,
\mathring{\tau}^{(d-1)/2}\nonumber \\ 
 &  & -L\frac{A_{d}}{d}\,\mathring{\tau}^{d/2}-\frac{\pi\,
   Q_{d+2,2}(4\mathring{\tau}L^{2})}{2^d\,
   \mathring{\tau}L^{d+1}}\bigg], \qquad\quad     
\end{eqnarray}
\begin{eqnarray}\label{eq:fpsi2spres}
f_{\psi,[2]}^{(\mathrm{sp},\mathrm{sp})}(L;\mathring{\tau})
 & = & \mathring{u}\,L\frac{n(n+2)}{4!}\bigg\{ \bigg[\frac{A_{d-1}}{2L}
 \mathring{\tau}^{(d-3)/2}\nonumber\\ &&\strut-A_{d}\,\mathring{\tau}^{(d-2)/2}
 +\frac{Q_{d,2}(4\mathring{\tau}L^{2})}{2^{d-1}\,\mathring{\tau}L^{d}}
 \bigg]^{2}\nonumber\\&&\strut +\frac{B_{d}}{L}
 \bigg[\frac{A_{2d-4}}{2L}\,\mathring{\tau}^{d-3}-A_{2d-3}\,
 \mathring{\tau}^{d-5/2}\nonumber \\ 
 &  & \strut +\frac{Q_{2d-3,2}(4\mathring{\tau}L^{2})}{2^{2d-4}
   \mathring{\tau}L^{2d-3}}\bigg]\bigg\},
\end{eqnarray}
respectively, where $f_{\psi,0}^{(\wp)}$ and $B_d$ are constants
defined in Eqs.~(\ref{eq:fpsi0wpapp}) and (\ref{eq:Bd}).

The contributions $-nL\,A_d\,\tau^{d/2}/d$ to $f_{\psi,[1]}^{(\wp)}$
have simple poles at $d=4$, which get cancelled upon renormalization
by the additive bulk counterterm $\propto \tau^2$ implied by the
subtraction~(\ref{eq:fbsubtr}). The two-loop terms
$f_{\psi,[2]}^{(\wp)}$ involve uv~singular bulk terms linear in $
A_d$ whose poles at $\epsilon=0$ get cancelled by the
$O(u)$~contribution to the counterterm $
(Z_\phi\,Z_\tau-1)\,\mu^2\tau\int_{\mathfrak{V}}\phi_R^2/2$. That no
pole-term singularities located at the boundary planes
$\mathfrak{B}_1$ and $\mathfrak{B}_2$ appear at $\epsilon=0$ in
$f_{\psi,[1]}^{(\mathrm{sp},\mathrm{sp})}$ and $f_{\psi,[2]}^{(\mathrm{sp},\mathrm{sp})}$ is
because both renormalized enhancement variables $c_1$ and $c_2$ are
zero. 

Since our main interest is in the renormalized residual free energy
$f_{\mathrm{res},R}^{(\wp)}$, we can avoid  dealing with additive
counterterms by focusing directly on its calculation. To determine its
non-zero mode contributions $f_{\psi,\mathrm{res},R}^{(\wp)}$, we must
subtract from the sums of the above one- and two-loop terms the bulk
and surface contributions and express the difference in terms of the
renormalized variables $\tau$ and $u$:
\begin{eqnarray}
  \label{eq:fxpsiwp}
\lefteqn{f_{\psi,\mathrm{res},R}^{(\wp)}(L;\tau,u,\mu)}&&\nonumber\\ &=&
  f_{\psi}^{(\wp)}(L;\mathring{\tau},\mathring{u})-L\,  f_b^{(\wp)}(\mathring{\tau},\mathring{u})-
  f_{\psi,s}^{(\wp)}(\mathring{\tau},\mathring{u} )\;.\qquad
\end{eqnarray}

From the results~(\ref{eq:fpsi1perres})--(\ref{eq:fpsi2spres}) one easily reads off the ($\wp$-independent)
bulk terms
\begin{equation}
  \label{eq:fbpsi}
  f_b=f_{b,0}-\frac{n\,A_d}{d}\,\mathring{\tau}^{d/2} +\mathring{u}\,\frac{n(n+2)}{4!}\,
  A_d^2\,\mathring{\tau}^{d-2}+O(\mathring{u}^2)
\end{equation}
as well as the $\wp$-dependent surface terms
\begin{eqnarray}
  \label{eq:fspsiper}
f_{\psi,s}^{\mathrm{(per)}}&=&f_{\psi,s,0}^{\mathrm{(per)}}+\frac{n\,A_{d-1}}{d-1}\,\mathring{\tau}^{(d-1)/2}\nonumber\\
&&\strut -\mathring{u}\,\frac{n(n+2)}{12}\,A_d A_{d-1}\,\mathring{\tau}^{d-5/2}+O(\mathring{u}^2)\qquad
\end{eqnarray}
and
\begin{equation}
  \label{eq:fspsispsp}
  f_{\psi,s}^{\mathrm{(sp,sp)}}=\frac{1}{2}\,f_{\psi,s}^{\mathrm{(per)}}-\mathring{u}\,\frac{n(n+2)}{4!}\,B_d\,A_{2d-3}\,
\mathring{\tau}^{d-5/2}+O(\mathring{u}^2).
\end{equation}

No confusion should arise from the fact that
$f_{\psi,s}^{\mathrm{(per)}}$ does not vanish. As is easily checked, and
our results for $f_\varphi^{\mathrm{(per)}}$ to be given below will show,
this term cancels exactly with the surface contribution to
$f_\varphi^{(\mathrm{per})}$, as it must. Of course, such cancellations are
neither expected nor occur for $\wp=(\mathrm{sp},\mathrm{sp})$ and  other
boundary conditions.

With the aid of the property
\begin{equation}\label{eq:diffeqQd2}
\frac{d}{dr}\bigg[\frac{Q_{d+2,2}(r)}{r}\bigg]
=-\frac{Q_{d,2}(r)}{4\pi r} 
\end{equation}
derived in appendix B of Ref.~\cite{DDG06}, the calculation of
$f^{(\wp)}_{\psi,\mathrm{res},R}$ becomes straightforward, giving
\begin{eqnarray}\label{eq:fpsiresRper}
 \frac{f^{(\mathrm{per})}_{\psi,\mathrm{res},R}}{\mu^{d-1}\,n}&=&
-\frac{4\pi\,Q_{d+2,2}(\mu^2\tau L^2)}{\tau\,(\mu
  L)^{d+1}}+\frac{u}{\mu L}\,
\frac{n+2}{4!} \nonumber\\ 
&&\times \Bigg\{\frac{1}{N_d}
\bigg[ A_{d-1}\,\tau^{(d-3)/2}
  +\frac{2\,Q_{d,2}(\mu^2\tau L^2)}{\tau\, (\mu L)^{d-1}}\bigg]^2
\nonumber\\ &&\strut
-\frac{\tau^{-\epsilon/2}-1}{\epsilon}\,
  \frac{8\,Q_{d,2}(\mu^2\tau L^2)}{(\mu L)^{d-1}}\Bigg\}+O(u^2)\;.
\nonumber\\ && 
\end{eqnarray}
and
\begin{eqnarray}
  \label{eq:fpsiresRsp}
\frac{f_{\psi,\mathrm{res},R}^{(\mathrm{sp},\mathrm{sp})}}{\mu^{d-1}n} & = &
-\frac{2\pi \,Q_{d+2,2}(4\mu^{2}\,\tau L^{2})}{\tau\,(2\mu L)^{d+1}}
+\frac{u}{\mu L}\frac{n+2}{4!}\nonumber \\ 
 &  & \times\Bigg\{ \frac{1}{N_{d}}
   \bigg[\frac{A_{d-1}}{2}\tau^{(d-3)/2} +\frac{Q_{d,2}(4\mu^{2}\tau L^{2})}{\tau(2\mu L)^{d-1}}\bigg]^{2}\nonumber \\
 &  & \strut +\frac{B_{d}}{N_{d}} \bigg[\frac{A_{2d-4}}{2}\tau^{d-3}
   +\frac{Q_{2d-3,2}(4\mu^{2}\tau \,L^{2})}{\tau\,(2\mu L)^{2d-4}}
 \bigg]\nonumber \\ 
 &  & \strut -\frac{\tau^{-\epsilon/2}-1}{\epsilon}
   \frac{2\,Q_{d,2}(4\mu^{2}\tau L^{2})}{(2\mu L)^{d-2}} \Bigg\}
 +O(u^{2})\;.\nonumber\\ &&
\end{eqnarray}

\subsection{Remaining free energy terms}\label{sec:fvarphi}

We next turn to the computation of $f_\varphi^{(\wp)}$. For
$\mathring{u}=0$ the Hamiltonian
$\mathcal{H}_{\mathrm{eff}}[\bm{\varphi}]$ describes a free field theory
whose two-point function is the familiar Gaussian bulk propagator
\begin{eqnarray}
  \label{eq:GGaussdmone}
  G_\infty^{(d-1)}(\bm{y}|\mathring{\tau})&=&\int_{\bm{p}}^{(d-1)}
    \,(p^2+\mathring{\tau})^{-1}\,e^{i\bm{p}\cdot\bm{y}} \nonumber\\ 
  &=&\frac{\left(\mathring{\tau}/y^2\right)^{(d-3)/4}}{(2\pi)^{(d-1)/2}}
  \,K_{(d-3)/2}\big(y\sqrt{\mathring{\tau}}\big) \qquad
\end{eqnarray}
in $d-1$ dimensions. As we have seen, using this as free propagator in
a Feynman graph expansion would lead to Feynman integrals that are
infrared divergent at $T_{c,\infty}$ and make the expansion
ill-defined beyond two-loop order. This suggests to work
with a free propagator whose mass parameter, firstly, remains positive
for $T\ge T_{c,\infty}$ when $L<\infty$, and secondly, has a
well-defined physical meaning beyond perturbation theory. A natural
candidate that has these properties is the inverse finite-size
susceptibility $r^{(\wp)}_L\equiv r_L$, defined by
\begin{equation}\label{eq:chiLdef}
  (r_L)^{-1}\,\delta_{\alpha\beta}=\chi_L\,\delta_{\alpha\beta}=\int
  d^{d-1}y\,\langle\varphi_\alpha(\bm{y})\,\varphi_\beta(\bm{0})\rangle^{\mathrm{cum}}\;.
\end{equation}
We therefore use 
\begin{equation}\label{eq:Gvarphifree}
  G_\varphi(\bm{y})\equiv
  G_{\infty}^{(d-1)}\big(\bm{y}|\mathring{r}_L\big)\;,  
\end{equation}
as free propagator. A tacit assumption underlying our calculation is
that the disordered phase is the correct reference state to expand
about for the parameter values of $L$ and $\tau \ge 0$ considered.
Since the transition temperature $T_{c,L}$ for finite $L$ is expected
to be lower than the bulk critical temperature $T_{c,\infty}$ (in
those cases of $d$ and $n$ for which a sharp transition occurs when
$L<\infty$), this is physically reasonable. However, there is no a
priori guarantee that extrapolations to $d=3$ of results based on
RG-improved perturbation theory will fulfill all necessary
requirements. In particular, we should check whether the so-obtained
approximate inverse finite-size susceptibilities $r^{(\wp)}_L$ remain
positive when $L<\infty$. This issue may be expected to be more
delicate for $\wp=(\mathrm{sp},\mathrm{sp})$ than for periodic boundary conditions.
The reason is that sp-sp boundary conditions are associated with a
multicritical point of the surface phase diagram (located at
$c=\tau=0$) at which the line of surface transition $T_{c,s}(c)$ meets
the bulk critical line (whose sections with $c>0$ and $c<0$ form the
lines of ordinary and extraordinary transitions, respectively)
\cite{Bin83,Die86a}. For finite $L$, one expects shifts of this
multicritical point and the phase boundaries. To account for these
shifts one would have to vary the surface enhancement variables $c_j$
as well, giving up the restriction $c_1=c_2=0$. This is a
difficult problem and beyond the scope of the present investigation. 

Let us represent the propagator~(\ref{eq:Gvarphifree}) by a red line,
the effective two-point vertex
$\mathring{\tau}-\mathring{r}_L-\sigma_\varphi$ by a red dot with two
legs, and the effective $k$-point vertices $\gamma^{(k)}$ with $k>2$
by red dots with $k$ legs. Then the Feynman graph expansion of
$f_\varphi^{(\wp)}(L)$ becomes
\begin{equation}\label{eq:Fgexpfvarphi}-f_\varphi^{(\wp)}(L)=
\,\,\raisebox{-7pt}{\includegraphics[width=20pt,clip]{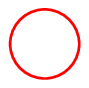}}
\,\,+\,\,
\raisebox{-7pt}{\includegraphics[width=20pt,clip]{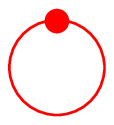}}
\,\,+\,\,
\raisebox{-12pt}{\includegraphics[width=16pt,clip]{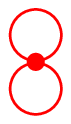}}
\,\,+\ldots\;.\phantom{\bigg[}
\end{equation}

The first graph on the right-hand side is given by
\begin{equation}\label{eq:fvarphi1}-\,\raisebox{-7pt}{\includegraphics[width=20pt,clip]{./1loop.eps}}=n\,f_{\varphi,0}+\frac{n}{2}
\int^{\mathring{r}_L}_{0}G^{(d-1)}_\infty(\bm{0}|t)\,dt 
\end{equation}
with
\begin{equation}\label{eq:fvarphi0}
f_{\varphi,0}=\frac{n}{2}\int_{\bm{p}}^{(d-1)}\ln p^{2}\;.
\end{equation}
Our results~(\ref{eq:gvarphi}) and (\ref{eq:sigphi})
for $g_\varphi^{-1}$ and $\sigma_\varphi$ imply that
\begin{eqnarray}
  \label{eq:twoptvert}
  \mathring{\tau}-\mathring{r}_L-\sigma_\varphi&=&
  \frac{n+2}{6}\,\frac{\mathring{u}}{L}\,
  G_{\infty}^{(d-1)}\big(\bm{0}|\mathring{r}_L\big)
  +O(\mathring{u}^2)\nonumber\\   &=&\frac{n+2}{6}\,
  \frac{A_{d-1}\mathring{u}}{L}\,
  \mathring{r}_L^{(d-3)/2} 
  +O(\mathring{u}^2)\;.\qquad \;\;\;
\end{eqnarray}
Using this in conjunction with the fact that the effective four-point
vertex, to first order in $\mathring{u}$, is a local $\varphi^4$
coupling with interaction constant $\mathring{u}/L$, one finds that
the contributions from the other two graphs can be written as
\begin{eqnarray}\label{eq:fvarphi2}\raisebox{-12pt}{\includegraphics[width=16pt,clip]{./2loop.eps}}
&=&-\frac{\mathring{u}}{L}\frac{n(n+2)}{4!}
\left[G_{\infty}^{(d-1)}(\bm{0}|\mathring{r}_{L})\right]^{2}
+O(\mathring{u}^2) \nonumber\\ &=&\strut
-\frac{1}{2}\,
\raisebox{-7pt}{\includegraphics[width=20pt,clip]{./1loopvert.eps}}\,\,
+O(\mathring{u}^2)\;.
\end{eqnarray}

Upon inserting  the $y\to 0$ limit of the free Gaussian
propagator~(\ref{eq:GGaussdmone}) into Eqs.~(\ref{eq:fvarphi1}) and
(\ref{eq:fvarphi2}), the required integrals can be performed to obtain
\begin{eqnarray}\label{eq:fvarphiwp}
f_{\varphi}^{(\wp)}
 & = &
 f_{\varphi,0}-n\,\frac{A_{d-1}}{d-1}\,\mathring{r}_{L}^{(d-1)/2}
 \nonumber\\ && 
 \strut -\frac{\mathring{u}}{L}\,\frac{n(n+2)}{4!}\,A_{d-1}^{2}\, 
 \mathring{r}_{L}^{d-3} +\ldots.
\end{eqnarray}

Just as the constants $f_{\psi,0}^{(\wp)}$ introduced in
Appendix~\ref{app:f10}, $f_{\varphi,0}$ involves uv divergent
contributions which are eliminated in the renormalized theory by the
additive renormalization of the free energy.
Furthermore, it should be remembered that
$\mathring{r}_L=\mathring{r}_L^{(\wp)}$ depends on the boundary
condition $\wp$. We have
\begin{equation}
  \label{eq:rLOuper}
\mathring{r}_L^{(\mathrm{per})}=\mathring{\tau}-
\frac{n+2}{6}\,\mathring{u}\bigg\{
A_d\,\mathring{\tau}^{(d-2)/2}  
-\frac{2\,Q_{d,2}(L^2\mathring{\tau})}{\mathring{\tau}\,L^d} \bigg\}
 +O(\mathring{u}^2)
\end{equation}
and
\begin{eqnarray} 
  \label{eq:rLOuspsp}
\mathring{r}_L^{(\mathrm{sp},\mathrm{sp})}&=&\mathring{\tau}-
\frac{n+2}{6}\,\mathring{u}\bigg\{
A_d\,\mathring{\tau}^{(d-2)/2}-\frac{2\,Q_{d,2}(4L^2\mathring{\tau})}{
  \mathring{\tau}\,(2L)^d} 
\nonumber\\ &&\strut
+\frac{A_{d-1}}{2L}\,\mathring{\tau}^{(d-3)/2}
 \bigg\}  +O(\mathring{u}^2) \;.
\end{eqnarray}
respectively.

The $O(u)$~contributions $\propto A_d=N_d/\epsilon$ of
$\mathring{r}_L^{(\wp)}$ have uv poles at
$\epsilon=0$. These are cured by the bulk counterterm
$(Z_\phi\,Z_\tau-1)\,\mu^2 \tau
\int_{\mathfrak{V}}\phi_R^2/2$. For the resulting  
renormalized dimensionless inverse susceptibilities
$r^{(\wp)}_L=Z_\phi\,\mathring{r}^{(\wp)}_L/\mu^2$ one obtains
\begin{eqnarray}
  \label{eq:rLperren}
  r^{(\mathrm{per})}_L
&=&\tau+\tau\,\frac{n+2}{6}\,\frac{u}{\tau^{\epsilon/2}}
  \bigg[\frac{\tau^{\epsilon/2}-1}{\epsilon/2}\nonumber\\ && \strut
     +\frac{2\,Q_{d,2}(\mu^2\tau L^2)}{(\mu^2
  L^2\,\tau)^{d/2}\,N_d}\bigg] +O(u^2)
\end{eqnarray}
and
\begin{eqnarray}
  \label{eq:rLspren}
  r^{(\mathrm{sp},\mathrm{sp})}_L
&=&\tau+\tau\,\frac{n+2}{6}\,\frac{u}{\tau^{\epsilon/2}}
  \bigg[\frac{\tau^{\epsilon/2}-1}{\epsilon/2}
 -\frac{A_{d-1}}{2\mu L\,\tau^{1/2}\,N_d} 
\nonumber\\ && \strut
     +\frac{2\,Q_{d,2}(4\mu^2\tau L^2)}{(4\mu^2
  L^2\,\tau)^{d/2}\,N_d}\bigg] +O(u^2)\;.
\end{eqnarray}

In Sec.~\ref{sec:Rscf} we will verify that these results comply with the
scaling form
\begin{equation}
  \label{eq:rLscf}
  r_L^{(\wp)}=r_\infty\,\mathsf{R}^{(\wp)}(L/\xi_\infty)\;,
\end{equation}
where 
\begin{equation}
  \label{eq:rinf}
  r_\infty=\chi_{\mathrm{b}+}^{-1}\,\tau^\gamma
\end{equation}
is the inverse bulk susceptibility,
and try to employ them to determine the scaling functions
$\mathsf{R}^{(\wp)}$ by means of the $\epsilon$ expansion.

Returning to the calculation of free energies, we now
subtract from $f_{\varphi}^{(\wp)}$ in Eq.~(\ref{eq:fvarphiwp}) the
surface contribution $f_{\varphi}^{(\wp)}|_{L=\infty,r_L=r_\infty}$ to
obtain the associated contributions $f_{\varphi,\mathrm{res}}^{(\wp)}$
to the residual free energies. Expressing the result in terms of
renormalized quantities then yields
\begin{eqnarray}\label{eq:fvarphiresRwp}
\frac{f^{(\wp)}_{\varphi,\mathrm{res},R}}{\mu^{d-1}n}&=&-\frac{A_{d-1}}{d-1}
\Big[r_L^{(d-1)/2}-r_\infty^{(d-1)/2}
\Big]\nonumber\\ && \strut -\frac{u}{\mu
  L}\frac{n+2}{4!}\,\frac{A_{d-1}^2}{N_d}
\,r_L^{d-3}+O(u^2)\qquad
\end{eqnarray}
for their renormalized analogs.

\subsection{General properties of the scaling functions}
\label{sec:genprop}

Before we embark on the calculation of the scaling functions
$\mathsf{R}^{(\wp)}(\mathsf{L})$ and $\Theta^{(\wp)}(\mathsf{L})$ of
the inverse finite-size susceptibility and the residual free energy,
it will be helpful to discuss some general properties they should
have.

In the limits $\mathsf{L}\to \infty$ and $\mathsf{L}\to 0$,
Eq.~(\ref{eq:rLscf}) must yield the correct bulk behavior and finite
positive finite-size susceptibility, respectively. This implies
\begin{equation}
  \label{eq:RasL}
  \mathsf{R}^{(\wp)}(\mathsf{L})\approx
  \begin{cases}1\,,&\text{for }\mathsf{L}\to \infty\;,\\[\medskipamount]
    \rho^{(\wp)}_{0+}\,
  \mathsf{L}^{\eta-2}\text{ with } \rho^{(\wp)}_{0+}>0\,,&\text{for }
  \mathsf{L}\to 0\;.
  \end{cases}
\end{equation}

Turning to the free-energy scaling functions
$\Theta^{(\wp)}(\mathsf{L})$, let us first consider their limiting
behavior as $\mathsf{L}\to 0$. This must comply with the requirement
that the finite-size free energy be analytic in $T$ at the bulk
critical temperature when $L<\infty$. As explained by KD, this
translates into the limiting form
\begin{eqnarray}\label{eq:smallLlimform}
 \Theta^{(\wp)}(\mathsf{L}) &\mathop{\approx}
  \limits_{\mathsf{L}\to 0}&
  \frac{a_{b+}\,\mathsf{L}^d}{\alpha(1-\alpha)(2-\alpha)}
  +\frac{a_{s+}^{(\wp)}
    \,\mathsf{L}^{d-1}}{\alpha_s(1-\alpha_s)(2-\alpha_s)}
\nonumber\\ &&\strut + \Delta_C^{(\wp)}
  +\sum_{k=1}^\infty\Delta_{k+}^{(\wp)}\,\mathsf{L}^{k/\nu} \;,
\end{eqnarray}
where $\alpha_s=\alpha+\nu$ is a familiar surface critical index (of
the surface excess specific heat \cite{Bin83,Die86a}).  Further,
$a_{b+}$ is a universal number whose $\epsilon$~expansion
\begin{equation}
  \label{eq:abplus}
  a_{b+}=\frac{n}{32\pi^2}\bigg\{1 +\frac{\epsilon}{2}\bigg[\ln(4\pi)
  -C_E+3\,\frac{n+2}{n+8}\bigg]+O(\epsilon^2)\bigg\} 
\end{equation}
may be gleaned from equation~(8.12) of Ref.~\cite{KD92a}. The plus
signs at $a_{b+}$, $a_{s+}$, and $\Delta_{k+}^{(\wp)}$ as
usual indicate that these  numbers pertain to the limit $\tau\to
0+$. 

The first two terms on the right-hand side of
Eq.~(\ref{eq:smallLlimform}) remove the singularities of the
subtracted bulk and surface contributions to
$f^{(\wp)}_{\mathrm{res},R}$; the remaining power series involves
integer powers of $\tau\propto (T-T_{c,\infty}) /T_{c,\infty}$. Note
that neither nonlinear contributions to the temperature scaling field
have been taken into account nor those of irrelevant bulk and surface
scaling fields. Both sources would entail corrections to the leading
thermal singularities of the bulk and surface free energies. The
implied additional terms nonanalytic in temperature would have to be
removed as well in the finite-size free energy and hence entail
further nonanalytic contributions to the limiting small-$\mathsf{L}$
form~(\ref{eq:smallLlimform}).

 The absence of boundaries in the case of periodic boundary
conditions implies that the surface amplitudes
$a_{s+}^{\mathrm{(per)}}$ are exactly zero. For the other case of
interest, $\wp=(\mathrm{sp},\mathrm{sp})$, one has
\begin{equation}
  \label{eq:asspsp}
   a^{\mathrm{(sp,sp)}}_{s+}=\frac{n}{128\pi}\bigg\{1
   +\epsilon\bigg[2+\ln\pi-C_E+\frac{n+2}{n+8}\bigg]+O(\epsilon^2)\bigg\} 
\end{equation}
according to KD's equations (E6) and (E9).

We next turn to a discussion of the limiting forms of the functions
$\Theta^{(\wp)}(\mathsf{L})$ as $\mathsf{L}\to\infty$. Since we have chosen
periodic boundary conditions along all $d-1$ parallel directions $y_j$,
no edge contributions $\sim L^{d-2}$ to the total free energy are
expected. Accordingly the residual free energy should decay
exponentially as $\mathsf{L}\equiv L/\xi_\infty\to\infty$. The asymptotic
behavior should simply follow from perturbation theory.

To become more precise, it is useful to recall the representations
(see, e.g., equations (4.2) and (4.12) of Ref.~\cite{Die86a})
\begin{equation}
  \label{eq:GperlargeL}
  G_L^{(\mathrm{per})}(\bm{x}_{12}|\mathring{\tau}) =\sum_{j=-\infty}^\infty
  G^{(d)}_\infty(\bm{x}_{12}+jL\,\bm{e}_z|\mathring{\tau}) 
\end{equation}
and
\begin{eqnarray}
  \label{eq:GsplargeL}
  G_L^{(\mathrm{sp},\mathrm{sp})}(\bm{x}_1,\bm{x}_2|\mathring{\tau}) &=&\sum_{j=-\infty}^\infty\big[
  G^{(d)}_\infty(\bm{x}_{12}+2jL\,\bm{e}_z|\mathring{\tau}) \nonumber\\ 
&&\strut +G^{(d)}_\infty\big(\bm{x}_{12}+2(jL+z_2)\,\bm{e}_z|\mathring{\tau}\big)\big]\nonumber\\
\end{eqnarray}
of the free propagators in terms of the bulk propagator
$G_\infty^{(d)}$, where
$\bm{x}_{12}=\bm{x}_1-\bm{x}_2=(\bm{y}_1,z_1)-(\bm{y}_2,z_2)$. 

The $j=0$ terms $G_\infty^{(d)}(\bm{x}_{12}|\mathring{\tau})$ yield the bulk
contributions of $G^{(\wp)}_L$. The $j=0$ term
$G_\infty^{(d)}(\bm{x}_{12}+2z_2|\mathring{\tau})$ and the $j=-1$ term
$G_\infty^{(d)}(\bm{x}_{12}+2(z_2-L)|\mathring{\tau})$ in
Eq.~(\ref{eq:GsplargeL}) represent surface contributions. Since
$G_\infty^{(d)}(\bm{x}|\mathring{\tau})$ decays exponentially as $|\bm{x}|\to\infty$
it is clear that of the remaining terms those involving spatial
differences that are constrained by the smallest lower bounds will
govern the limiting large-$\mathsf{L}$ behavior of the functions
$\Theta^{(\wp)}$. In the case of periodic boundary conditions, this
applies to the $j=\pm 1$ terms, which involve position vectors of
lengths $\ge \mathsf{L}$.  Hence $\Theta^{(\mathrm{per})}(\mathsf{L})$
must vary as $\sim e^{-\mathsf{L}}$ in the limit
$\mathsf{L}\to\infty$, up to powers of $\mathsf{L}$.

On the other hand, for $\wp=(\mathrm{sp},\mathrm{sp})$, there are four
contributions involving position vectors constrained by the 
lower-distance bound $2L$ which govern the large-$\mathsf{L}$
behavior. Thus  $\Theta^{(\mathrm{sp},\mathrm{sp})}(\mathsf{L})$
must decay  $\sim e^{-2\mathsf{L}}$, up to powers of $\mathsf{L}$. 

To elaborate on these arguments, one can employ the above expressions
(\ref{eq:GperlargeL}) and (\ref{eq:GsplargeL}) for the free
propagators in perturbation theory, dropping all of their
summands that do not contribute to the leading large-$\mathsf{L}$
behavior. In the case of the one-loop integrals it is again convenient to
first determine the large-$\mathsf{L}$ forms of their
$\mathring{\tau}$-derivatives and then integrate with respect to
$\mathring{\tau}$. However, from our perturbative results gathered in
Eqs.~(\ref{eq:fpsi1perres})--(\ref{eq:fpsi2spres}),
(\ref{eq:fvarphi1}), (\ref{eq:fvarphi2}), and (\ref{eq:fvarphiresRwp}),
the one- and two-loop Feynman integrals with \emph{all} contributions
to the free propagators included can be inferred. Thus no renewed
calculation is necessary. To determine the large-$\mathsf{L}$ behavior
of the $\Theta^{(\wp)}$ we must merely replace the functions $Q_{d,2}$
and $Q_{d+2,2}$ by their asymptotic forms~(\ref{eq:Qd2largeas}) given
in Appendix~\ref{app:Qd2}.  This yields
\begin{equation}
  \label{eq:Thetaperlargeas}
  \Theta^{\mathrm{(per)}}(\mathsf{L}) \mathop{\approx}
  \limits_{\mathsf{L}\to\infty}- \frac{n}{(2\pi)^{(d-1)/2}}\,
  \mathsf{L}^{(d-1)/2}\,e^{-\mathsf{L}}\, [1+O(u^*)]
\end{equation}
and
\begin{equation}
  \label{eq:Thetasplargeas}
  \Theta^{\mathrm{(sp,sp)}}(\mathsf{L}) \mathop{\approx}
  \limits_{\mathsf{L}\to\infty}- \frac{n}{2^{d}\,\pi^{(d-1)/2}}\,
  \mathsf{L}^{(d-1)/2}\,e^{-2\mathsf{L}}\, [1+O(u^*)]\;.
\end{equation}

Finally, let us briefly recall what can be said about the behavior of
the $\tau<0$ analogs of the scaling functions $\Theta^{(\wp)}$, which
we denote as $\Theta^{(\wp)}_-(\mathsf{L})$, at the transition
temperature $T_{c,L}$ of the film in those cases where a sharp
transition to a long-range ordered phase is possible for finite $L$,
such as in the Ising case $n=1$ for bulk dimension $d=3$.  As a
function of the temperature deviation $t_L=(T-T_{c,L})/T_{c,L}$, the
excess free energy per cross-sectional area $A$ must have a
contribution that behaves as $\sim t_L^{2-\alpha_{d-1}}$ as $t_L\to
0$, where $\alpha_{d-1}$ is the specific heat exponent for bulk
dimension $d-1$. The transition point translates into a nonzero value
$\mathsf{L}_0$ at which the functions $\Theta^{(\wp)}_-(\mathsf{L})$ behave in
a nonanalytic fashion. Standard matching of the temperature
singularities then yields the behavior
\begin{equation}
  \label{eq:ThetaasTcL}
  \Theta^{(\wp)}_-(\mathsf{L})\sim \big|\mathsf{L}^{1/\nu}
  -\mathsf{L}_0^{1/\nu}\big|^{2-\alpha_{d-1}}\;, 
\end{equation}
where $\nu\equiv\nu_d$, as before, is the correlation-length
exponent of the $d$-dimensional bulk system.

A well-known consequence is that the critical-temperature shift varies
as \cite{Fis71,Bar83}
\begin{equation}
  \label{eq:crittempshift}
  (T_{c,\infty}-T_{c,L})/T_{c,\infty} \sim L^{-1/\nu}\;.
\end{equation}
This conclusion that the shift exponent is given by $1/\nu$ is
more or less automatic when the finite-size scaling
form~(\ref{eq:scffintro}) of the residual free energy applies and
hence is in complete accordance with our theory.

\subsection{Scaling functions of inverse finite-size susceptibilities}
\label{sec:Rscf}

We proceed by combining our perturbative results of
Sec.~\ref{sec:fpsi} and \ref{sec:fvarphi} with the RG to compute the
desired scaling functions, beginning with those of the inverse
finite-size susceptibilities $r_L^{(\wp)}$. To this end we use the RG
flow to map the original renormalized theory to one corresponding to
the choice $\ell=(\mu\,\xi_\infty)^{-1}$ of the scale parameter.  The
running coupling constant $\bar{u}(\ell)$ can be replaced by the
fixed-point value $u^*=3\epsilon /(n+8)+O(\epsilon^2)$ at the expense of
neglecting corrections to scaling $\sim \bar{u}(\ell)-u^*$. The
running temperature variable $\bar{\tau}(1/\mu\,\xi_\infty)$ is
exactly unity (at the required first order in $u^*$, when $\tau >0$).

As straightforward consequences of Eqs.~(\ref{eq:rLperren}) and
(\ref{eq:rLspren}) we thus obtain
\begin{equation}
  \label{eq:Rper}
  \mathsf{R}^{(\mathrm{per})}(\mathsf{L})=1+\epsilon\,\frac{n+2}{n+8}\,
\frac{16\pi^2\, Q_{4,2}(\mathsf{L}^2)}{\mathsf{L}^4}+o(\epsilon)
\end{equation}
and
\begin{equation}
  \label{eq:Rsp}
  \mathsf{R}^{(\mathrm{sp},\mathrm{sp})}(\mathsf{L})=1+\epsilon\,\frac{n+2}{n+8}\,
\frac{\pi^2\,
  Q_{4,2}(4\mathsf{L}^2)-\pi\,\mathsf{L}^3}{\mathsf{L}^4}
+o(\epsilon)\;.
\end{equation}

Using the asymptotic forms (\ref{Q42smallr}) and (\ref{Q42larger}) of
$Q_{4,2}(4\mathsf{L}^2)$ for small and large $\mathsf{L}$, one sees
that these results are in conformity with the limiting
behavior~(\ref{eq:RasL}). The amplitudes $\rho_{0+}^{(\wp)}$ are found
to be 
\begin{equation}
  \label{eq:rho0}
  \rho_{0+}^{\mathrm{(per)}}=4\,\rho_{0+}^{\mathrm{(sp,sp)}}+o(\epsilon)
  =\epsilon\frac{n+2}{n+8}\,\frac{2\pi^2}{3}+o(\epsilon).
\end{equation}

The approach to the large-$\mathsf{L}$ limit
$\mathsf{R}^{(\wp)}(\infty)=1$ is qualitatively different for periodic
and sp-sp boundary conditions: it is of an exponential and algebraic
form in the first and latter cases, respectively. 

\begin{figure}[htb]
  \centering
\includegraphics[width=0.85\columnwidth,clip]{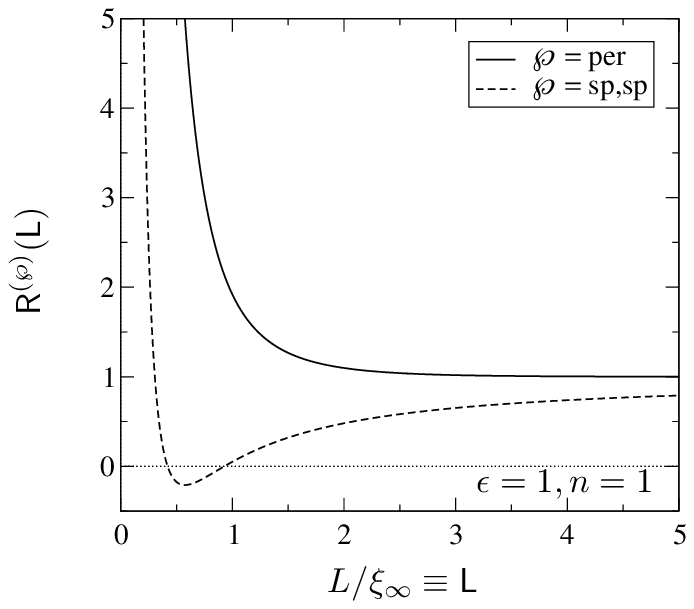}
  \caption{Plots of the scaling functions 
    $\mathsf{R}^{(\mathrm{per})}(\mathsf{L})$ and
    $\mathsf{R}^{(\mathrm{sp},\mathrm{sp})}(\mathsf{L})$ for $n=1$ and $d=3$, as
    obtained from Eqs.~(\ref{eq:Rper}) and (\ref{eq:Rsp}) by setting $\epsilon=n=1$.}\label{fig:R}
\end{figure}

In Fig.~\ref{fig:R} we have plotted the extrapolations to $d=3$ of the
$O(\epsilon)$ results~(\ref{eq:Rper}) and (\ref{eq:Rsp}), obtained by
setting $\epsilon=1$, for the one-component case $n=1$. It reveals
another important difference: The extrapolation
$\mathsf{R}^{\mathrm{(per)}}(\mathsf{L})|_{\epsilon=n=1}$ remains
positive for all $\mathsf{L}>0$, reassuring us thus that the theory is
consistent in that the disordered state about which we expanded
satisfies this necessary stability condition. By contrast, the
extrapolation
$\mathsf{R}^{\mathrm{(sp,sp)}}(\mathsf{L})|_{\epsilon=n=1}$
\emph{becomes negative} for $0.42\lesssim\mathsf{L}\lesssim 0.93$.
When extrapolated to $d=3$ in this na{\"\i}ve manner, the theory thus
yields a violation of stability of the disordered state in this range
of parameters. 

\begin{figure}[htb]
  \centering
\includegraphics[width=0.85\columnwidth,clip]{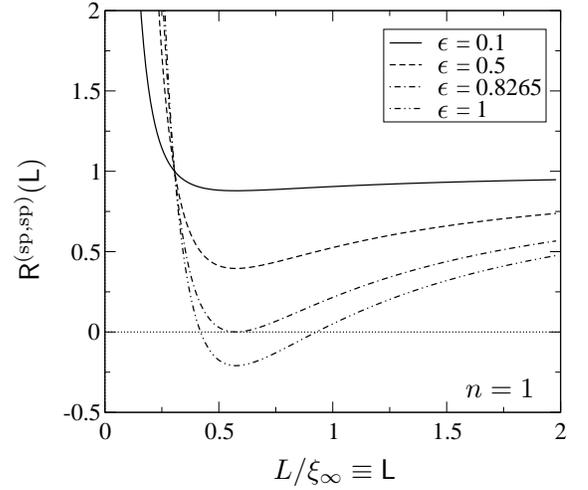}
\caption{Plots of $O(\epsilon)$ result for the scaling function
  $\mathsf{R}^{(\mathrm{sp},\mathrm{sp})}(\mathsf{L})$ for $n=1$ and the
  indicated values $\epsilon=0.1$, $0.5$, $0.8265$, and
  $1$.}\label{fig:Rspvareps}
\end{figure}

It is to be emphasized that this is a problem already for KD's
original extrapolations of their $\epsilon$-expansion results for the
Casimir effect. As we shall see below, in our reformulated field
theory it will show up in an even more exposed fashion. Note, however,
that negative values of the $O(\epsilon)$ result for
$\mathsf{R}^{\mathrm{(sp,sp)}}(\mathsf{L})$ are encountered only for
values of $\epsilon\gtrsim 0.8265$. This is illustrated in
Fig.~\ref{fig:Rspvareps}, where
$\mathsf{R}^{\mathrm{(sp,sp)}}(\mathsf{L})|_{n=1}$ is plotted for
several different values of $\epsilon$.
It is conceivable, although not at all guaranteed, that extrapolations
based on perturbative calculations to higher orders will yield
positive definite functions $\mathsf{R}^{(\mathrm{sp},\mathrm{sp})}(\mathsf{L})$.
As already remarked above, we believe that in systematic studies of
the stability of the disordered phase, besides temperature, the
surface enhancement variables $c_1$ and $c_2$ should be allowed to
vary  --- a difficult task, which is beyond the scope of
our present analysis.

\subsection{Scaling functions of  the residual free energies}
\label{sec:fresscf}

To determine the free-energy scaling functions 
$\Theta^{(\wp)}(\mathsf{L})$, we start with the decompositions
\begin{equation}
  \label{eq:freswpR}
  f_{\mathrm{res},R}^{(\wp)}(L;\tau,u,\mu)=
  f_{\psi,\mathrm{res},R}^{(\wp)}
  +f_{\varphi,\mathrm{res},R}^{(\wp)}
\end{equation}
and 
\begin{equation}\label{eq:Thetawpdec}
  \Theta^{(\wp)}(\mathsf{L})=\Theta_\psi^{(\wp)}(\mathsf{L})
  +\Theta_\varphi^{(\wp)}(\mathsf{L})\;.
\end{equation}
analogous to Eq.~(\ref{eq:fLfpsiFvarphi}). We now substitute the
perturbative expressions~(\ref{eq:fpsiresRper}),
(\ref{eq:fpsiresRsp}), and (\ref{eq:fvarphiresRwp}) for
$f_{\psi,\mathrm{res},R}^{(\wp)}$ and
$f_{\varphi,\mathrm{res},R}^{(\wp)}$, insert Eqs.~(\ref{eq:rLperren})
and (\ref{eq:rLspren}) for $r_L^{(\mathrm{per})}$ and
$r_L^{(\mathrm{sp},\mathrm{sp})}$, together with their common large-$L$ limit
\begin{equation}
  \label{eq:rinfty}
  r_\infty=\tau+\tau\,\frac{n+2}{6}\,\frac{u}{\tau^{\epsilon/2}}
  \frac{\tau^{\epsilon/2}-1}{\epsilon/2}+O(u^2)
\end{equation}
for the inverse bulk susceptibility $r_\infty$.  This yields the
(truncated) series-expansion results for
$f_{\mathrm{res},R}^{(\wp)}(L;\tau,u,\mu)$ on which our subsequent
analysis is based. We now combine them with the RG, proceeding along
the lines explained and followed above. 

The functions $\Theta_\psi^{(\wp)}$ have conventional expansions in
integer powers of $\epsilon$, which to first order in $\epsilon$
follow directly from Eqs.~(\ref{eq:fpsiresRper}) and
(\ref{eq:fpsiresRsp}). Our results are
\begin{eqnarray}
  \label{eq:Thetapsiper}
   \Theta_\psi^{(\mathrm{per})}(\mathsf{L})&=&-n\,\frac{4 \pi \,[
    Q_{6,2}\left(\mathsf{L}^2\right)
    -\epsilon\,R_{6,2}(\mathsf{L}^2)]}{\mathsf{L}^2} \nonumber\\ &&
  \strut +n\epsilon\,\frac{n+2}{n+8}\,  \frac{\left[\mathsf{L}^3+8\,
      \pi\,Q_{4,2}\left(\mathsf{L}^2\right)\right]^2}{8\mathsf{L}^4}
   \nonumber\\ && \strut
+O(\epsilon^2)
\end{eqnarray}
and
\begin{eqnarray}
  \label{eq:Thetapsisp}
   \lefteqn{\Theta_\psi^{(\mathrm{sp},\mathrm{sp})}(\mathsf{L})}&&\nonumber\\ &=&
   -n\,\pi\,\frac{(1+\epsilon\ln
   2)\,Q_{6,2}(4\mathsf{L}^2) 
-\epsilon\,R_{6,2}(4\mathsf{L}^2)}{16\,\mathsf{L}^2}
\nonumber \\ &&\strut  
+n\,\epsilon\,\frac{n+2}{n+8}\,
\frac{\pi\,Q_{4,2}(4\mathsf{L}^2)\big[2\mathsf{L}^3+\pi\,Q_{4,2}(4\mathsf{L}^2)\big]}{32\,\mathsf{L}^4} 
\nonumber \\ &&\strut +O(\epsilon^2)\;,
\end{eqnarray}
where $R_{6,2}$ is defined by
\begin{equation}
R_{d,\sigma}(r)\equiv\frac{\partial Q_{d,\sigma}(r)}{\partial d}\;.
\end{equation}

Inspection of KD's work reveals that the non-zero mode part of their
$O(\epsilon)$ expression for $\Theta^{(\mathrm{sp},\mathrm{sp})}$ coincides with
their result for $\Theta^{(D,D)}$. By consistency, the latter should
agree with our result~(\ref{eq:Thetapsisp}) for
$\Theta_\psi^{\mathrm{(sp,sp)}}$. This is indeed the case, as can easily
be verified by comparison, using the relation
\begin{eqnarray}
  \label{eq:Rdgab}
  R_{6,2}(r)&=&\frac{r^3}{32\pi^3}\Big[\Big(C_E
  -\frac{8}{3}+\ln\frac{r}{\pi}\Big)\, g_{3/2,0}\big(\sqrt{r}/2\big)
  \nonumber\\ &&\strut +g_{3/2,1}\big(\sqrt{r}/2\big)\Big] 
\end{eqnarray}
implied by Eq.~(\ref{eq:Qd2ga}).

A consistency check can also be made for $\Theta_\psi^{\mathrm{(per)}}$ by
noting that the contribution produced by the $\mathsf{L}^3$~term
in $[\ldots]^2$ of Eq.~(\ref{eq:Thetapsiper}) corresponds to the
subtracted $k_0=0$ part. Thus, by dropping it, we should recover
KD's result  for $\Theta^{\mathrm{(per)}}$ given in the third line of
their equations (6.13). Confirming this is again straightforward by
virtue of Eq.~(\ref{eq:Rdgab}). 

We stress that unlike the full scaling functions
$\Theta^{(\wp)}(\mathsf{L})$, their non-zero mode parts
$\Theta_\psi^{(\wp)}(\mathsf{L})$ do \emph{not in general decay
  exponentially} as $\mathsf{L}\to\infty$ and \emph{should not be
  expected} to have this property. This is because the zero-mode
pieces projected out involve contributions to the residual free energy
density $f_{\mathrm{res},R}$ that decay as $1/L$. These imply
contributions to $\Theta_\psi^{(\wp)}(\mathsf{L})$ that vary as
$\mathsf{L}^{d-2}$ in the large-$\mathsf{L}$ limit. Inspection of our
result~(\ref{eq:Thetapsiper}) shows that the $O(\epsilon)$ term of
$\Theta_\psi^{\mathrm{(per)}}$ indeed grows as $\mathsf{L}^2$. By
contrast, our $O(\epsilon)$ result~(\ref{eq:Thetapsisp}) for
$\Theta_\psi^{(\mathrm{sp},\mathrm{sp})}$ is seen to decay exponentially for
large $\mathsf{L}$ because both the functions $Q_{d,2}$ and $R_{d,2}$
do so [cf.\ Eq.~(\ref{eq:Qd2largeas})]. The absence of an analogous
$O(\epsilon)$ contribution $\sim \mathsf{L}^2$ to
$\Theta_\psi^{(\mathrm{sp},\mathrm{sp})}$ is due to the cancellation of the two
terms of $f_{\psi,[2]}$ in Eq.~(\ref{eq:fpsi2spres}) proportional to
$A_{d-1}^2/L$ and $B_d/L$, respectively. Of course, if such
cancellation did not occur then the above-mentioned equality of
$\Theta_\psi^{(\mathrm{sp},\mathrm{sp})}$ with $\Theta_\psi^{(D,D)}$ to first
order in $\epsilon$ would be impossible.

We next turn to the computation of the functions
$\Theta^{(\wp)}_\varphi$. This is a considerably more subtle problem,
which requires care. It should be clear that we must not simply expand
in powers of $\epsilon$. The small-$\mathsf{L}$ behavior of the
scaling functions $\Theta^{(\wp)}$ should be compatible with the
behavior found for $\tau=0$ in Ref.~\cite{DGS06} and hence yield the
contributions $\sim\epsilon^{3/2}$ to the Casimir amplitudes. The
mechanism by which this happens is that the inverse susceptibilities
$r^{(\wp)}_L(\tau,u^*)$ approach nonzero limits $r_L^*(0,u^*)=O(u^*)$
as $\tau\to 0$ when $L<\infty$. The $O(\epsilon^{3/2})$ terms then result
from the contributions $\sim r_L^{(d-1)/2}=r_L^{3/2+O(\epsilon)}$ to
$f^{(\wp)}_{\varphi,\mathrm{res},R}$ in Eq.~(\ref{eq:fvarphiresRwp}).

On the other hand, if we expand in powers of $\epsilon$, taking
$\mathsf{L}$ (i.e., $\tau$) to be positive, then KD's series-expansion
results to order $\epsilon$ still ought to be recovered.

Substitution of the respective one-loop results~(\ref{eq:rLperren}) and
(\ref{eq:rLspren}) for
$r_L^{(\wp)}$ in the zero-mode free-energy
contribution~(\ref{eq:fvarphiresRwp}), in conjunction with
Eq.~(\ref{eq:Thetafresrel}), yields
\begin{equation}
  \label{eq:Thetavarphiwp}
  \Theta_\varphi^{(\wp)}(\mathsf{L})=\frac{n\,\mathsf{L}^3}{12\pi}
    \bigg\{1 -\frac{3\pi}{2\mathsf{L}}\,\frac{n+2}{n+8}\,\epsilon
-[\mathsf{R}^{(\wp)}(\mathsf{L})]^{3/2}\bigg\}\;,
\end{equation}
where $\mathsf{R}^{(\wp)}(\mathsf{L})$ represents the respective $O(\epsilon)$
expression for these scaling functions
given in Eqs.~(\ref{eq:Rper}) and (\ref{eq:Rsp}).

In the case of periodic boundary conditions, which we consider first,
the combination of Eqs.~(\ref{eq:Rper}), 
(\ref{eq:Thetawpdec}), (\ref{eq:Thetapsiper}), 
and (\ref{eq:Thetavarphiwp}) leads to
\begin{eqnarray}
  \label{eq:Thetaperfinres}
\lefteqn{
\Theta^{(\mathrm{per})}(\mathsf{L})
}&&\nonumber\\
&=&-n\,\frac{4 \pi \big[Q_{6,2}(\mathsf{L}^2)
-\epsilon\,R_{6,2}(\mathsf{L}^2)\big]}{\mathsf{L}^2}
\nonumber\\ &&\strut
 +n\,\epsilon\,\frac{n+2}{n+8}\,  \frac{{[\mathsf{L}^3 +8
      \pi\,Q_{4,2}(\mathsf{L}^2)]}^2
    -\mathsf{L}^6}{8\mathsf{L}^4}  
\nonumber\\ &&\strut 
+\frac{n\,\mathsf{L}^3}{12\pi}\Bigg[1- \bigg(1+
\epsilon\,\frac{n+2}{n+8}\,
\frac{16\pi^2\, Q_{4,2}(\mathsf{L}^2)}{\mathsf{L}^4}\bigg)^{3/2}
\Bigg]\,.\nonumber\\ &&\strut+o\big(\epsilon^{3/2}\big)
\end{eqnarray}

This result has the following properties:

(i) Upon expanding it to first order in $\epsilon$ [
i.e., the term $[\ldots]^{3/2}$ in Eq.~(\ref{eq:Thetaperfinres})]
when $\mathsf{L}=0$, one recovers KD's result.

(ii) The limiting value $\Theta^{\mathrm{(per)}}(0)$ agrees with our
$O(\epsilon^{3/2})$ result for
\begin{eqnarray}
  \label{eq:Deltaper}
  \Delta_C^{(\mathrm{per})}
 &=&  -\frac{n\pi^{2}}{90}
  +\frac{n\pi^{2}\epsilon}{180}\bigg[1-C_E-\ln \pi
    +\frac{2\,\zeta^{\prime}(4)}{\zeta(4)}
\nonumber\\ &&\strut
     +\frac{5}{2}\frac{n+2}{n+8}
   \bigg]
 -\frac{n\pi^{2}}{9\sqrt{6}} \left
   (\frac{n+2}{n+8}\right)^{3/2}\epsilon^{3/2}\nonumber\\ &&\strut
+O(\epsilon^2) 
\end{eqnarray}
in Ref.~\cite{DGS06}.

(iii) The small-$\mathsf{L}$ behavior of
$\Theta^{\mathrm{(per)}}(\mathsf{L})$ differs from the requested one
specified in Eq.~(\ref{eq:smallLlimform}) by terms $\propto
\epsilon^{3/2}\,\mathsf{L}$; we have
\begin{equation}
  \label{eq:smallLviolat}
  \Theta^{\mathrm{(per)}}(\mathsf{L})\mathop{\approx}
  \limits_{\mathsf{L}\to 0}  \Delta_{C}^{(\mathrm{per})}
  +\frac{n\,\pi}{2\sqrt{6}}
  \bigg(\frac{n+2}{n+8}\bigg)^{3/2}\,\epsilon^{3/2}\,\mathsf{L}
  +O(\mathsf{L}^{2})\;.    
\end{equation}
In KD's result the term linear in $\mathsf{L}$ that is at variance
with the limiting form~(\ref{eq:smallLlimform}) is of first order in
$\epsilon$;  here it is of the same order
$\epsilon^{3/2}$ to which we determined $\Delta_C^{\mathrm{(per)}}$. 
 
(iv) The large-$\mathsf{L}$ asymptotic behavior of
$\Theta^{\mathrm{(per)}}(\mathsf{L})$ is in conformity with
Eq.~(\ref{eq:Thetaperlargeas}), just as KD's result is.

It is gratifying that our result has the properties (i), (ii), and
(iv). On the other hand, it still does not fully comply with the
small-$\mathsf{L}$ form~(\ref{eq:smallLlimform}) dictated by the
analyticity of the total finite-size free energy at $T_{c,\infty}$,
though the violations now occur at the corresponding higher order
$\epsilon^{3/2}$. 

\begin{figure}[htb]
  \centering
  \includegraphics[clip,width=0.85\columnwidth]{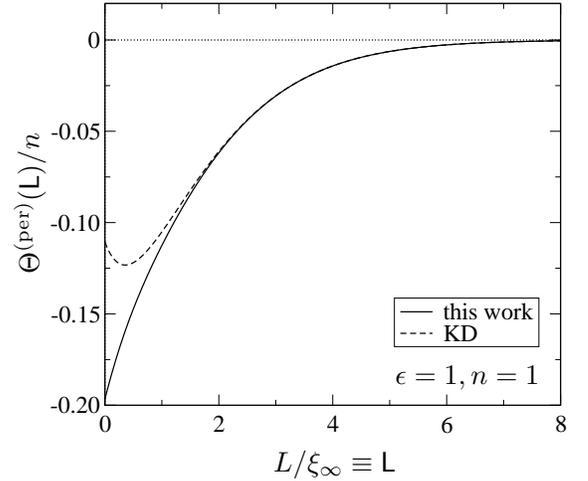}
  \caption{Extrapolations to $d=3$ of the scaling function
    $\Theta^{\mathrm{(per)}}(\mathsf{L})$ for $n=1$, obtained
    by setting $\epsilon=1$ in Eq.~(\ref{eq:Thetaperfinres}) and KD's
    original $O(\epsilon)$ result, respectively.}
  \label{fig:Thetapern1d3}
\end{figure}

In Fig.~\ref{fig:Thetapern1d3} our result for the scaling function
$\Theta^{\mathrm{(per)}}(\mathsf{L})$ with $n=1$ and $d=3$, obtained by
setting $\epsilon=1$ in Eq.~(\ref{eq:Thetaperfinres}), is plotted and
compared with its analog for KD's $\epsilon$-expansion result. The
minimum in KD's extrapolation result appears to be due to the
inadequate handling of the zero-mode contributions. Our extrapolation
gives a monotonic behavior at small $\mathsf{L}$, which agrees better
both with the Monte Carlo data of Ref.~\cite{DK04} as well as with
improved, more recent ones \cite{Huc:tbp,VGMD07}.

\begin{figure}[htb]
  \centering
  \includegraphics[clip,width=0.85\columnwidth]{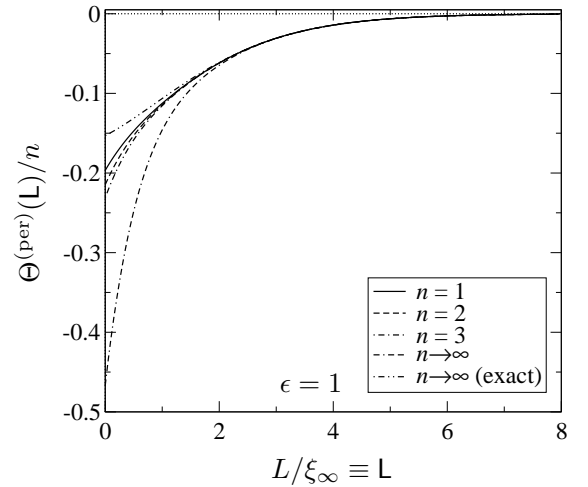}
  \caption{Extrapolations to $d=3$ of the scaling function
    $\Theta^{\mathrm{(per)}}(\mathsf{L})$ for $n=2$, $n=3$, and
    $n=\infty$, obtained by setting $\epsilon=1$ in
    Eq.~(\ref{eq:Thetaperfinres}). For comparison, the
    spherical-model result for $d=3$ \cite{Dan96,BDT00}, which is
    exact for $n=\infty$, is also shown.} \label{fig:Thetapern123inf}
\end{figure}

In Fig.~\ref{fig:Thetapern123inf} analogous extrapolations to $d=3$ of
the scaling functions for $n=2$, $n=3$, and $n=\infty$ are displayed,
along with the exact spherical-model result for $d=3$. The comparison
with the extrapolations based on KD's $O(\epsilon)$ results displayed
in Fig.~\ref{fig:Thetaextr} indicates, on the one hand, that the
extrapolations for given $n$ oscillate as the order of the series
expansion is increased and, on the other hand, that the variations
with order are the bigger the larger $n$ is.  

Next, we consider the case of sp-sp boundary conditions. In discussing
extrapolations to $d=3$ dimensions, we shall restrict ourselves to
the $n=1$ component case. The reason should be clear: Only when $n=1$
is a multicritical point expected to occur at $T_{c,\infty}$ and a
finite enhancement of the surface interaction constants
\cite{rem:surfaniso,DE82,DE84}.

A first problem was encountered in our investigation of the inverse
finite-size susceptibility $r_L^{(\mathrm{sp},\mathrm{sp})}$: Our one-loop result
for the scaling function
$\mathsf{R}^{(\mathrm{sp},\mathrm{sp})}(\mathsf{L},n{=}1,\epsilon{=}1)$ becomes
negative for $0.42\lesssim\mathsf{L}\lesssim 0.93$.  Clearly,
convincing predictions for the scaling functions
$\Theta^{\mathrm{(sp,sp)}}(\mathsf{L})$ in $d=3$ dimensions must also
fulfill necessary stability conditions such as the positive
definiteness of $r_L$. Thus the violation of this stability criterion
of the disordered state is a problem even for extrapolations to $d=3$
of KD's original $O(\epsilon)$ results. In our result given by the
combination of Eqs.~(\ref{eq:Rsp}), (\ref{eq:Thetawpdec}),
(\ref{eq:Thetapsisp}), and (\ref{eq:Thetavarphiwp}) it manifests
itself in an obvious, striking manner: For values of the scaling
variable $\mathsf{L}$ in the mentioned interval, the extrapolation to
$d=3$ would yield complex numbers for
$\Theta^{\mathrm{(sp,sp)}}(\mathsf{L})$. 

A further problem occurs for large $\mathsf{L}$: The contribution to
$\Theta^{(\mathrm{sp},\mathrm{sp})}(\mathsf{L})$ originating from the
term $\propto 1/\mathsf{L}$ in curly brackets in
Eq.~(\ref{eq:Thetavarphiwp}) in conjunction with the part $\propto
1/\mathsf{L}$ of $\mathsf{R}^{\mathrm{(sp,sp)}}$ produce a
large-$\mathsf{L}$ behavior of the form
$O(\epsilon^2)\,\mathsf{L}+O(\epsilon^3)\,\mathsf{L}^0$. Thus, unless
we subtract these asymptotic terms $\propto \epsilon^2\,\mathsf{L}$ and
$\propto \epsilon^3\,\mathsf{L}^0$, our approximation for
$\Theta^{(\mathrm{sp},\mathrm{sp})}(\mathsf{L})$ will not have a
finite limit as $\mathsf{L}\to \infty$, and hence yield unacceptable
results at $d=3$ even in the regime $\mathrm{L}\gtrsim 0.92$ where the
positivity condition $\mathsf{R}^{(\mathrm{sp},\mathrm{sp})}>0$ is
satisfied.

The combination of these two problems puts us in a bad position to
suggest convincing extrapolations to $d=3$. Let us, however, note some
appealing properties the result given by Eqs.~(\ref{eq:Rsp}),
(\ref{eq:Thetawpdec}), (\ref{eq:Thetapsisp}), and
(\ref{eq:Thetavarphiwp}) has. All above properties (i)--(iii)
of the small-$\mathsf{L}$ behavior hold just as in the case of periodic
boundary conditions. That is, KD's $O(\epsilon)$ results are recovered
when the term
$[\mathsf{R}^{(\mathrm{sp},\mathrm{sp})}(\mathsf{L})]^{3/2}$ in
Eq.~(\ref{eq:Thetavarphiwp}) is expanded in powers of $\epsilon$.
Second, the limiting value $\Theta^{(\mathrm{sp},\mathrm{sp})}(0)$
reproduces the expansion of the Casimir amplitude to order
$\epsilon^{3/2}$,
\begin{eqnarray}\label{eq:Delspsp}
\Delta_C^{(\mathrm{sp},\mathrm{sp})} &=& -\frac{n\pi^{2}}{1440}
+\frac{n\pi^{2}\epsilon}{2880}\bigg[1-C_E
-\ln(4\pi)
+\frac{2\zeta^{\prime}(4)}{\zeta(4)}\nonumber\\ 
&&\strut +\frac{5}{2}\frac{n+2}{n+8}\bigg] 
 -\frac{n\pi^{2}}{72\sqrt{6}} \bigg(\frac{n+2}{n+8}
 \bigg)^{3/2}\epsilon^{3/2}
\nonumber \\ && \strut
+ O(\epsilon^{2})\;.
\end{eqnarray}
Third, the term linear in $\mathsf{L}$ that violates the limiting
form~(\ref{eq:smallLlimform}) is of order $\epsilon^{3/2}$ rather than
linear in $\epsilon$. We have
\begin{equation}
  \label{eq:ThetaspsmallL}
  \Theta^{(\mathrm{sp},\mathrm{sp})}(\mathsf{L}) \mathop{\approx}
  \limits _{\mathsf{L}\to0} \Delta_{C}^{(\text{sp,sp})}
  +\frac{n\,\pi}{4\sqrt{6}}
  \bigg(\frac{n+2}{n+8}\bigg)^{3/2}\epsilon^{3/2}\,
  \mathsf{L}+O(\mathsf{L}^{2}). 
\end{equation}

Furthermore, the large-$\mathsf{L}$ behavior still is in accordance
with Eq.~(\ref{eq:Thetasplargeas}) in the sense that the differences
are of higher than first order in $\epsilon$. However, as already
mentioned, it would lead to extrapolations to $d=3$ that grow
$\sim\mathsf{L}$ in the limit $\mathsf{L}\to\infty$ unless
contributions of the form $\sim O(\epsilon^2)\,
\mathsf{L}+O(\epsilon^3)\,\mathsf{L}^0$ are subtracted.

\begin{figure}[htb]
  \centering
\includegraphics[width=0.87\columnwidth,clip]{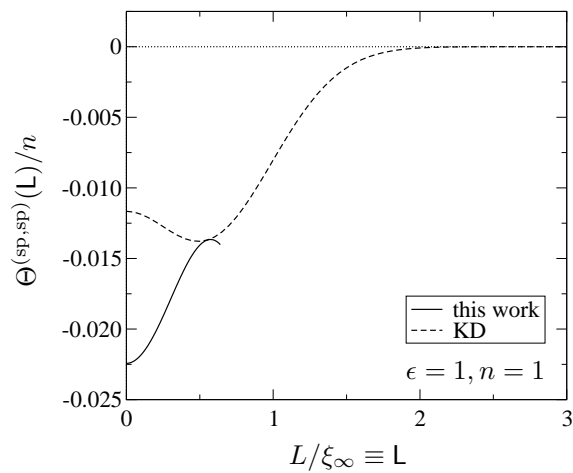}
\caption{Extrapolations to $d=3$ of the scaling function
  $\Theta^{(\mathrm{sp},\mathrm{sp})}$ for $n=1$. The curve labelled
  ``KD'' corresponds to the $O(\epsilon)$ results of Ref.~\cite{KD92a}
   evaluated at $\epsilon=1$; the other one is our result given by
  Eqs.~(\ref{eq:Rsp}), (\ref{eq:Thetawpdec}), (\ref{eq:Thetapsisp}),
  and (\ref{eq:Thetavarphiwp}), with $\epsilon$ set to one. For
  further explanations, see main text.}
  \label{fig:Thetaspn1d3}
\end{figure}

In Fig.~\ref{fig:Thetaspn1d3} we have plotted the extrapolated scaling
function $\Theta^{(\mathrm{sp},\mathrm{sp})}(\mathsf{L})$ one obtains
from Eqs.~(\ref{eq:Rsp}), (\ref{eq:Thetawpdec}),
(\ref{eq:Thetapsisp}), and (\ref{eq:Thetavarphiwp}) upon setting
$\epsilon=1$, together with its analog (labeled KD) implied by the
$\epsilon$-expansion result. The former function is depicted only for
values $\mathsf{L}$ below the lower threshold $\simeq 0.42$ beyond
which the extrapolated scaling function
$\mathsf{R}^{(\mathrm{sp},\mathrm{sp})} $ of the inverse
susceptibility becomes negative. We have refrained from displaying it
(or appropriate modifications of it) for values larger than the upper
positivity threshold $\simeq 0.93$.  In view of the $O(\epsilon^2)$
corrections the result would require for large $\mathsf{L}$ to ensure
its decay for $\mathsf{L}\to \infty$, we have no convincing reasons to
expect such ad hoc modifications to yield much better results in this
regime of $\mathsf{L}$ than the extrapolated $\epsilon$ expansion.

One might wonder whether the above problems could be avoided by a
different choice of the free propagator $G_\varphi$ in
Eq.~(\ref{eq:Gvarphifree}). For example, one might want to use one
whose mass parameter is simply the sum of the free contribution
$\mathring{\tau}$ and the first-order perturbative
correction~(\ref{eq:sigphi}). We have in fact explored this
possibility. It yields a modified scaling function
$\tilde{\Theta}^{(\mathrm{sp},\mathrm{sp})}(\mathsf{L})$ whose
large-$\mathsf{L}$ behavior  must be corrected by $O(\epsilon^2)$ contributions
to avoid unacceptable divergences. Once this is done, its
extrapolation to $\epsilon=1$ gives real values for all $\mathsf{L}$.
We refrain from displaying the results because we consider them
unsatisfactory for two reasons. First of all, as explained before
Eq.~(\ref{eq:Gvarphifree}), we believe that the use of the inverse
finite-size susceptibility $r_L^{-1}$ as mass parameter is the more
natural choice. Second, the fact that one is able to produce a
well-defined extrapolated scaling function
$\tilde{\Theta}^{(\mathrm{sp},\mathrm{sp})}(\mathsf{L})$ does not cure
the problem that the $O(\epsilon)$ result for the scaling function
$\mathsf{R}^{(\mathrm{sp},\mathrm{sp})}(\mathsf{L})$ of $r_L$ becomes
negative when extrapolated to $\epsilon=1$. Convincing improvments
should yield meaningful extrapolation results for both the scaling
function $\Theta^{(\mathrm{sp},\mathrm{sp})}(\mathsf{L})$ and $r_L$
within one and the same consistent approximation scheme.
Evidently, further work is necessary to improve on the present
unsatisfactory state of these results for sp-sp boundary conditions.

On the other hand, the behavior of our results at small $\mathsf{L}$
may be expected to be  superior to those based on the $\epsilon$-expansion. 
One indication is that, in the case of periodic boundary
conditions, our results are in conformity with the exact solution in
the large-$n$ limit (see Sec.~\ref{sec:compsmres}). 

\section{Comparison with spherical-model results for periodic boundary
  conditions} \label{sec:compsmres}

As is well known, for translation invariant systems results that are
exact in the limit $n\to\infty$ can be obtained from the exact
solution of spherical models \cite{Sta68}. The self-consistent
equations from which the scaling function
$\Theta_{\mathrm{SM}}^{\mathrm{(per)}}(\mathsf{L})$ for the spherical
model with periodic boundary conditions must be determined can be
found in the literature \cite{Dan96,Dan98,BDT00,DDG06}. Our aim here
is to verify the consistency of our results for periodic boundary
conditions with the exact solution of the spherical model for
$2<d=4-\epsilon<4$. Making an analogous check for $\wp=(\mathrm{sp},\mathrm{sp})$
is a much harder challenge and will not be attempted here. The reason
is that the presence of surfaces in general destroys translation
invariance perpendicular to the boundary planes. The large-$n$ limit
of $n$-vector models on slabs with two parallel boundary planes
$\mathfrak{B}_1$ and $\mathfrak{B}_2$ is known to correspond to a
modified spherical model involving separate constraints on the sums
$\sum_{j\in \text{ layer }z}S_j^2$ of the squares of the spin
variables for each layer $z$ \cite{Kno73}. The resulting
self-consistent equations, while not difficult to determine, involve a
$z$-dependent self-consistent pair interaction and so far have not
been solved analytically.

The exact solution for the spherical-model scaling function
$\Theta_{\mathrm{SM}}^{\mathrm{(per)}}(\mathsf{L})=\lim_{n\to\infty}\Theta^{(\mathrm{per})}(\mathsf{L})/n
$ may be gleaned from
Ref.~\cite{DDG06}, where this function was denoted as $Y_0$. It is
given by
\begin{eqnarray}\label{eq:Thetasmstart}
  \Theta_{\mathrm{SM}}^{\mathrm{(per)}}(\mathsf{L})&=&\frac{A_d}{2}\,\mathsf{L}^{d-2}\,(\mathsf{R}_0-\mathsf{L}^2)
  -\frac{A_d}{d}\big(\mathsf{R}_0^{d/2}-\mathsf{L}^{d}\big)\nonumber\\
&&\strut
  -\frac{4\pi\,Q_{d+2,2}(\mathsf{R}_0)}{\mathsf{R}_0} \;,
\end{eqnarray}
where $\mathsf{R}_0=\mathsf{L}^2\,
\mathsf{R}_{\mathrm{SM}}^{\mathrm{(per)}}(\mathsf{L})$ 
is a solution to
\begin{equation}
  \frac{2\,Q_{d,2}(\mathsf{R}_0)}{\mathsf{R}_0}=A_d\,
  \big(\mathsf{R}_0^{(d-2)/2} -\mathsf{L}^{d-2}\big) \;.
\end{equation}

The latter equation is easily solved for small $\epsilon$. Since $A_d$
has a pole $\propto \epsilon^{-1}$, the left-hand side starts to
contribute at $O(\epsilon)$. One obtains
\begin{equation}\label{eq:R0}
  \mathsf{R}_0(\mathsf{L})=\mathsf{L}^2 +\epsilon\,
  \frac{16\pi^2\,Q_{4,2}(\mathsf{L}^2)}{\mathsf{L}^2}+o(\epsilon) \;,
\end{equation}
which becomes
\begin{equation}
  \mathsf{R}_0(0)=\epsilon \,\frac{2}{3}\,\pi^2+o(\epsilon)
\end{equation}
at the bulk critical point. Comparison of these results with ours for
$\mathsf{L}^2\,\mathsf{R}_{\mathrm{SM}}^{\mathrm{(per)}}(\mathsf{L})$
contained in  Eqs.~(\ref{eq:Rper}) and (\ref{eq:rho0}) shows that the
latter reduce to them in the limit $n\to\infty$.

Turning to $\Theta_{\mathrm{SM}}^{\mathrm{(per)}}(\mathsf{L})$, we note
that according to the representation (\ref{eq:Qdsiggps}) of
$Q_{d+2,2}$, two contributions in Eq.~(\ref{eq:Thetasmstart}) can be
combined as
\begin{eqnarray}
  \label{eq:Qd2rewr}
  \lefteqn{ \frac{-A_d}{d}\,\mathsf{R}_0^{d/2}-\frac{4\pi\,
      Q_{d+2,2}(\mathsf{R}_0)}{\mathsf{R}_0}  }&&\nonumber\\
  &=&\frac{-A_{d-1}}{d-1}\,
  \mathsf{R}_0^{(d-1)/2} -\sum_{k=0}^\infty
  \frac{a_k(d)}{k!}\,(-\mathsf{R}_0)^k \;, 
\end{eqnarray}
where
\begin{equation}
  \label{eq:ak}
  a_k(d)=\frac{\pi^{(d-1)/2}}{(2\pi)^{2
      k}}\,\Gamma[k+(1-d)/2]\,\zeta(1-d+2k)\;.
\end{equation}
Except for $a_2(d)$, which has a simple pole at $d=4$, the
coefficients $a_k(d)$ are regular at $d=4$. We therefore separate the
contribution from the first term in the second line of
Eq.~(\ref{eq:Qd2rewr})
\begin{eqnarray}
  \label{eq:Adm1term}
  \lefteqn{\frac{-A_{d-1}}{d-1}\,
  \mathsf{R}_0^{(d-1)/2}}&&\nonumber\\ &=&\frac{-1}{12\pi}\left[\mathsf{L}^2 +\epsilon\,
  \frac{16\pi^2\,Q_{4,2}(\mathsf{L}^2)}{\mathsf{L}^2}\right]^{3/2}
+o\big(\epsilon^{3/2}\big)\;,
\end{eqnarray}
where we substituted  $\mathsf{R}_0$ by its expansion~(\ref{eq:R0}),
and then expand the remaining contributions to
$\Theta_{\mathrm{SM}}^{\mathrm{(per)}}(\mathsf{L})$ in powers of
$\epsilon$. This gives
\begin{eqnarray}
  \label{eq:ThetaSMepsexp}
  \Theta_{\mathrm{SM}}^{\mathrm{(per)}}(\mathsf{L})&=&\frac{-1}{12\pi}
  \left[\mathsf{L}^2 +\epsilon\, 
  \frac{16\pi^2\,Q_{4,2}(\mathsf{L}^2)}{\mathsf{L}^2}\right]^{3/2}
+\frac{\mathsf{L}^3 }{12\pi}\nonumber\\ &&\strut
-\frac{4\pi[Q_{6,2}(\mathsf{L}^2)
  -\epsilon\,R_{6,2}(\mathsf{L}^2)]}{\mathsf{L}^2} \nonumber\\
&&\strut
+\epsilon\, \frac{8\pi^2\,Q_{4,2}(\mathsf{L}^2)}{\mathsf{L}^2}\bigg[
\frac{\mathsf{L}}{4\pi}
+\frac{Q_{4,2}(\mathsf{L}^2)}{\mathsf{L}^2}\bigg]\nonumber\\ &&\strut
+o\big(\epsilon^{3/2}\big) \;.
\end{eqnarray}
The result agrees with the one for
$\Theta^{(\mathrm{per})}(\mathsf{L})/n$ given in
Eq.~(\ref{eq:Thetaperfinres}) if the factor $(n+2)/(n+8)$ is replaced
by its large-$n$ limit ($=1$). In particular, its value 
at $\mathsf{L}=0$,
\begin{eqnarray}
  \Theta_{\mathrm{SM}}^{\mathrm{(per)}}(0)&=&-\frac{\pi^2}{90}
  +\frac{\pi^2}{180} \left[\frac{7}{2}-C_E-\ln\pi
    +\frac{2\zeta'(4)}{\zeta(4)}\right]\epsilon \nonumber\\ &&\strut
  -\frac{\pi^2}{9\sqrt{6}}\,\epsilon^{3/2}+O(\epsilon^{2})\;, 
\end{eqnarray}
coincides with the limit $\lim_{n\to\infty}\Delta_C^{(\mathrm{per})}/n$ of
the expansion~(\ref{eq:Deltaper}). The same holds for the coefficient
of the term linear
in $\mathsf{L}$, for which we find
\begin{equation}
  \label{eq:Thetaprime}
  \frac{d\Theta_{\mathrm{SM}}^{\mathrm{(per)}}(\mathsf{L})}{
    d\mathsf{L}} \bigg|_{\mathsf{L}=0}=
  \frac{\pi}{2\sqrt{6}}\,\epsilon^{3/2} +o\big(\epsilon^{3/2}\big) 
\end{equation}
which is consistent with Eq.~(\ref{eq:smallLviolat}).

\section{Summary and concluding remarks}
\label{sec:concl}

In this paper we have reconsidered the use of renormalized field
theory near the upper critical bulk dimension $d^*=4$ to the study of
finite-size scaling in slabs of finite thickness and the thermodynamic
Casimir effect. In previous work \cite{DGS06} it had become clear that
in those cases where the boundary conditions involve zero modes in
Landau theory at the bulk critical point, the conventional RG-improved
perturbation theory based on the $\epsilon$ expansion becomes
ill-defined at $T_{c,\infty}$ due to infrared singularities. This
could be remedied by means of a reorganization of field theory, which
revealed that noninteger powers such as $\epsilon^{3/2}$ appear in the
small-$\epsilon$ expansion.

Our main aim here was to examine how the calculation of scaling
functions describing the large length-scale behavior of the residual
free energy and the Casimir force near the bulk critical point can be
reconciled with these findings, so that the results of
Ref.~\cite{DGS06} for $T=T_{c,\infty}$ are recovered in the
appropriate limit. 

We were able to show that consistent scaling functions can indeed be
obtained both for the case of periodic and sp-sp boundary
conditions. It became clear that the ill-definedness of the
conventional $\epsilon$-expansion theory due
to zero modes manifests itself already at two-loop order inasmuch as
contributions found at this order were found to have no power-series
expansion in $\epsilon$ at $T_{c,\infty}$ since they  vary
$\sim \epsilon^{3/2}$.

In calculations of crossover scaling functions by means of RG-improved
perturbation theory near an upper critical dimension one usually is
faced with the following problem. The RG commonly achieves the proper
exponentiation of the infrared singularities only at the unstable
fixed point. However, it does not normally do this --- at least, not
automatically --- for the modified singularities that occur as the
scaled crossover variable becomes large. Knowledge about the corresponding
asymptotic behavior frequently is obtained from other sources, such as
RG analyses of a different model or fixed point, or short-distance
expansion. Representative examples are the calculation of the
two-point correlation function \cite{AF74}, the crossover at a
bicritical point \cite{Hor76,AG78}, and the crossover from critical to
Goldstone-mode behavior in isotropic ferromagnets \cite{SH78,Law81}.
To obtain and verify the correct singularities of the behavior to
which the crossover occurs by means of the $\epsilon$ expansion, it
must be supplemented by appropriate assumptions, or preferably
knowledge, about the respective asymptotic forms. In some cases it has
even been possible to design RG procedures that yield the correct
asymptotic behaviors at both the unstable fixed point as well as the
stable one to which the crossover occurs \cite{AG78,Law81}, albeit
with somewhat limited range of applicability and success.

Similar problems evidently had to be expected in the study of the
problems considered here --- finite-size effects and thermodynamic
Casimir forces. However, the challenges are actually greater and the
difficulties more severe. Ideally, one would like to have a theory
that has the power to correctly treat the infrared singularities at
both the bulk critical point as well as the film critical point and
moreover is capable of handling the corresponding dimensional crossover.
For reasons discussed at the end of Sec.~\ref{sec:revftapp}, such
ambitious goals would be unrealistic for a theory based on an
expansion about the upper critical dimension. We therefore set out to
reach more modest goals, namely: to modify and correct the previous
theory by an appropriate treatment of the zero mode in such a way
that
(i) RG-improved perturbation theory becomes well-defined for
temperatures $T\ge T_{c,\infty}$, 
{(ii)} reasonable scaling functions result whose
limiting behavior complies with the theory's predictions directly at
$T_{c,\infty}$ and can be extrapolated to $d=3$ dimensions, and 
(iii) hence bring it into a state comparable to the one it has for the
non-zero-mode boundary conditions $\wp=\mathrm{ap}$, $(D,D)$, and
$(D,\mathrm{sp})$.

We feel that, on the whole, our results are encouraging, in
particular, for the case of periodic boundary conditions, where
besides achieving (i)--(iii), we were able to demonstrate consistency
with the exact large-$n$ solution. Moreover, the scaling function
obtained by extrapolation to $d=3$, at least in the one-component
case, appears to agree reasonably well with Monte Carlo results
\cite{Huc:tbp,VGMD07}. 

The case of sp-sp boundary conditions turned out to be more delicate.
First of all, we found that the one-loop expression for the scaling
function $\mathsf{R}^{(\mathrm{sp},\mathrm{sp})}(\mathsf{L})$ of the inverse
finite-size susceptibility becomes negative in a small regime of
$\mathsf{L}=L/\xi_\infty$ when $\epsilon$ exceeds the value $\simeq
0.8265$ (see Figs.~\ref{fig:R} and \ref{fig:Rspvareps}). This tells us that all extrapolations of free-energy scaling
functions and Casimir forces to $d=3$ based on approximations which
yield the same one-loop scaling function
$\mathsf{R}^{(\mathrm{sp},\mathrm{sp})}(\mathsf{L})$ are questionable, at least
in the regime where the positivity condition
$\mathsf{R}^{(\mathrm{sp},\mathrm{sp})}(\mathsf{L})\ge 0$ is violated. This
applies both to KD's original extrapolation and ours (see
Fig.~\ref{fig:Thetaspn1d3}). 

Our investigation of this case also revealed another problem:
Perturbative RG calculations do not necessarily yield the correct
asymptotic large-$\mathsf{L}$ behavior, at least not automatically.
This applies even for the conventional $\epsilon$ expansion in cases
where no zero mode is present inasmuch as the algebraic prefactors
$\sim\mathsf{L}^{(d-1)/2}$ appearing in the asymptotic exponential
behaviors of $\Theta^{(\mathrm{per})}(\mathsf{L})$ and
$\Theta^{(\mathrm{sp},\mathrm{sp})}(\mathsf{L})$ given in
Eqs.~(\ref{eq:Thetaperlargeas}) and (\ref{eq:Thetasplargeas}),
respectively, are obtained only in $\epsilon$-expanded form. However,
it is more troublesome in the cases studied here, especially, for
$\wp=(\mathrm{sp},\mathrm{sp})$. The reason may be understood as follows. On the
one hand, we encountered powers of inverse finite-size
susceptibilities we had to retain to ensure consistency with the
behavior at $T_{c,\infty}$. On the other hand, by expanding other
contributions in $\epsilon$, $\mathsf{L}$-dependent terms of order
$\epsilon^2$ and higher are dropped which may be needed to cancel
similar $\mathsf{L}$-dependent contributions originating from the
unexpanded powers of $\mathsf{R}$ in order to avoid incorrect or even
divergent large-$\mathsf{L}$ behavior of the scaling functions.

A qualitative difference between periodic and sp-sp boundary
conditions is that the latter involve, even in the semi-infinite case
$L=\infty$, both $d$- and $(d-1)$-dimensional critical behavior, rather
than just a dimensional crossover.  With hindsight it is therefore
perhaps not too surprising that the latter turned out to be the more
difficult case.

As remarked earlier, special surface transitions are expected to occur
in three bulk dimensions only in the $n=1$ case. When $n>1$,
anisotropic special transitions should be possible if the continuous
$O(n)$ symmetry is broken by an appropriate easy-axis spin anisotropy
at the surface \cite{rem:surfaniso,DE82,DE84}. This is because surface
phases with long-range order should not be thermodynamically stable at
temperatures $T> T_{c,\infty}$, by analogy with the Mermin-Wagner
theorem \cite{MW66}. However, the $O(2)$ case is exceptional in that a
surface phase with quasi-long-range order should be possible. In fact,
recent Monte Carlo work \cite{PL90,DBN05} indicated that the surface
phase transition is of Kosterlitz-Thouless type. Thus 
a multicritical surface-bulk point at which the line of these surface
transitions reaches $T_{c,\infty}$ should exist as well
\cite{Die86a}, and was reported to be found in the cited Monte Carlo
analyses. 

Since the lambda transition of Helium involves a (real-valued)
two-component order parameter, this $O(2)$ case is of potential
relevance for Casimir forces in confined liquid He. In the case of
$^3$He-$^4$He mixtures in contact with a substrate (see, e.g.,
\cite{BI05,MD06}), $^4$He usually gets enriched near the wall and a
superfluid surface film may form there. Since order-parameter
correlations decay algebraically in it, the bulk transition in the
presence of such a critical surface phase is reminiscent of the
special transition. Whether the central issue we were concerned with
in this work --- the presence of zero modes in Landau theory ---
arises also in the study of the thermodynamic Casimir effect in such
systems and what its consequences are remains to be seen. A proper
analysis of this question requires generalizations of our model. To
describe mixtures, a second density besides the order parameter is
needed. In addition, care must be taken to ensure a proper description
of the Kosterlitz-Thouless-like surface transition.

The present work suggests extensions and complementary work along
several lines. The situation in the case of sp-sp boundary conditions
is rather unsatisfactory. To improve it, it would be desirable to
extend our analysis by allowing the surface enhancement variables
$c_j$ to vary. In fact, in order to clarify the effects of finite size
on the phase diagram, and in turn resolve the issue in which range of
parameters the disordered state is thermodynamically stable, such a
generalization appears to be unavoidable. An appealing other aspect of
it would be that by varying the $c_j$, one could smoothly interpolate between
the boundary conditions $\wp=(D,D)$, $(D,\mathrm{sp})$, and
$(\mathrm{sp},\mathrm{sp})$.

In view of the great technical and conceptual difficulties one is
faced with in such analytical approaches, we believe that careful
checks of their predictions by alternative means such as Monte Carlo
simulations are absolutely necessary. For a long time detailed studies
of the thermodynamic Casimir effect by this method existed only for
the case of periodic boundary conditions \cite{KL96,DK04,Kre94}.
However, recently new simulation strategies for investigating this
effect in lattice spin systems with free boundary conditions have been
developed \cite{Huc07,Huc:tbp,VGMD07}.  As a result, systematic
numerical studies of the thermodynamic Casimir effect under all sorts
of interesting boundary conditions have become possible.

On the side of analytical theories, it would be interesting to explore
whether the present approach can be combined with existing RG
approaches at fixed dimension $d$ for the study of bulk and surface
critical phenomena \cite{Par80,SD89,DS94,DS98}. Another important
challenge is to develop reliable analytical approaches by which the
Casimir effect can be investigated below the bulk and film critical
temperatures. Recent investigations of the ordered phase based on
Landau theory or RG-improved Landau theory \cite{MGD07,ZSRKC07}
certainly should not remain the final word since they fail to give
correct descriptions of the critical behavior at both the bulk
critical point as well as at eventual film critical points. In
addition, they are known to be sometimes even qualitatively wrong
inasmuch as they may predict phases with long-range order that can 
be shown to be destroyed by thermal fluctuations.

\acknowledgments

We are indebted to Daniel Dantchev for calling our attention to the
problem with the $n$-dependence of the $\epsilon$-expansion results of
Ref.~\cite{KD92a}, which initiated our interest in this work. It is
our pleasure to also thank him and Mykola Shpot for stimulating
discussions. We owe thanks to Alfred Hucht for informing us about his
Monte Carlo simulations prior to publication, and to Andrea Gambassi
for sending us a preprint of Ref.~\cite{VGMD07}.

Finally, we gratefully acknowledge partial support by the
Deutsche Forschungsgemeinschaft under Grant No.~Di-378/5.

\appendix

\section{Computation of required integrals}\label{app:reqintegr}

The one- and two-loop Feynman integrals for $f^{(\wp)}_\psi$ involve
the free $\psi$-propagator~(\ref{eq:psiprop}) at coincident points
$\bm{x}=\bm{x}^\prime$. Substitution of the eigenfunctions~(\ref{eq:per}) and
(\ref{eq:sp}) into Eq.~(\ref{eq:psiprop}) yields
\begin{equation}
G_{L,\psi}^{(\mathrm{per})}(\bm{x};\bm{x}|\mathring{\tau})
=\frac{2}{L}\int_{\bm{p}}^{(d-1)}\sum_{m=1}^{\infty}\frac{1}{p^{2}
  +(2\pi m/L)^{2}+\mathring{\tau}}\label{eq:freepropagatorPBC}
\end{equation}
and
\begin{equation}
G_{L,\psi}^{(\mathrm{sp},\mathrm{sp})}(\bm{x};\bm{x}|\mathring{\tau})
=\frac{2}{L}\int_{\bm{p}}^{(d-1)}\sum_{m=1}^{\infty}
\frac{\cos^{2}(\pi mz/L)}{p^{2}+(\pi m/L)^{2}+\mathring{\tau}}\;.
\label{eq:freepropagatorspsp}
\end{equation}

In order to compute the integrals $I^{(\wp)}_{1}(L;\mathring{\tau})$
introduced in Eq.~(\ref{eq:Iwpj}), we add and subtract a summand with
$m=0$ and then use Poisson's summation formula (see, e.g.,
Eq.~(4.8.28) in \cite{MF53}) 
\begin{equation}
\sum_{m=-\infty}^{\infty}f(am)=
\sum_{j=-\infty}^{\infty}\int_{-\infty}^{\infty}\frac{dt}{a}f(t)e^{2\pi
  ijt/a}\;.
\label{eq:PoissonSumFormula}
\end{equation}
Recalling the definition~(\ref{eq:definitionQdsigma})  of the
functions $Q_{d,\sigma}$ and performing the required momentum
integrals, one arrives at the results for
$I_{1}^{\mathrm{(per)}}(L;\mathring{\tau})$ and
$I_{1}^{\mathrm{(sp,sp)}}(L;\mathring{\tau})$ given in
Eqs.~(\ref{eq:I1per}) and (\ref{eq:I1sp}). 

Turning to the calculation of $I_2^{(\wp)}(L;\mathring{\tau})$, we
note that for $\wp=\mathrm{per}$ we have
\begin{equation}\label{eq:I2perres}
I_{2}^{\mathrm{(per)}}(L;\mathring{\tau})=
\big[I_{1}^{\mathrm{(per)}}(L;\mathring{\tau})\big]^{2}
\end{equation}
as a consequence of translation invariance along the $z$~direction.

To compute $I_{2}^{(\mathrm{sp},\mathrm{sp})}$, we use the representation
\begin{eqnarray}\label{eq:I2sprep}
I_{2}^{(\mathrm{sp},\mathrm{sp})}(L;\mathring{\tau}) &=& 
\int_{0}^{L} \frac{dz}{L}\left[
  G_{L,\psi}^{(\mathrm{sp},\mathrm{sp})}(\bm{x}; \bm{x}|\mathring{\tau})
\right]^{2}\nonumber\\ 
 & = & \sum_{m,m'=1}^{\infty}
\int_{\bm{p}_{1}}^{(d-1)}
 \int_{\bm{p}_{2}}^{(d-1)}\frac{4}{L^{2}}\int_{0}^{L}
 \frac{dz}{L}\nonumber\\ &&\times
 \frac{\cos^{2}(k_{m}z)
   \cos^{2}(k_{m'}z)}{(p_{1}^{2}+k_{m}^{2}
   +\mathring{\tau})(p_{2}^{2}+k_{m'}^{2} +\mathring{\tau})}
\end{eqnarray} 
and the fact that
\begin{equation}\label{eq:cos2cos2}
4\int_{0}^{L}\frac{dz}{L}\cos^{2}(k_{m}z)\cos^{2}(k_{m'}z)
=1 +\frac{1}{2}\,\delta_{mm'}
\end{equation}
for $k_m=m\pi/L$ and $m,m'\ne 0$.
Upon inserting the latter result into Eq.~(\ref{eq:I2sprep}), one
arrives at
\begin{eqnarray}
I_{2}^{(\mathrm{sp},\mathrm{sp})}(L;\mathring{\tau}) &=&\big[
  I_{1}^{(\mathrm{sp},\mathrm{sp})}(L;\mathring{\tau})\big]^2 
\nonumber\\ && \strut
+
\frac{1}{2\,L^2}\,A_{d-1}^2\sum_{m=1}^{\infty}\left(k_{m}^{2}
  +\mathring{\tau}\right)^{d-3},\qquad 
\end{eqnarray} 
where $A_d$ was defined in Eq.~(\ref{eq:Ad}).

In order to evaluate the second term on the right-hand side of this
equation, we employ the analytical continuation of the Epstein-Hurwitz
zeta function discussed in appendix A of Ref.~\cite{ER89}, namely
\begin{eqnarray}\label{eq:EpsteinHurwitzZeta}
\lefteqn{\sum_{j=1}^{\infty}\left(j^{2}+\alpha^{2}\right)^{-s}}&&\nonumber\\
&=&\strut
-\frac{1}{2\,\alpha^{2s}}
+\frac{\sqrt{\pi}}{2\,\alpha^{2s-1}\,\Gamma(s)}
\bigg[\Gamma(s-1/2)+\frac{4}{(\pi\alpha)^{(1-2s)/2}}\nonumber\\ && \times
   \sum_{m=1}^{\infty}
  m^{s-1/2}\,K_{(1-2s)/2}(2\pi m\alpha)\bigg],\quad
s\ne 1/2\;. 
\end{eqnarray}
A straightforward calculation then yields
\begin{eqnarray}
  \label{eq:I2spspres}
  I^{(\mathrm{sp},\mathrm{sp})}_2(L;\mathring{\tau})& = &
  \big[I^{(\mathrm{sp},\mathrm{sp})}_1(L;\mathring{\tau})\big]^{2} \nonumber\\
  &&\strut +\frac{B_{d}}{L}\bigg[\frac{A_{2d-4}}{2L}\, \mathring{\tau}^{d-3} 
  -A_{2d-3}\,\mathring{\tau}^{d-5/2}\nonumber \\
  &  & \strut +\frac{Q_{2d-3,2}(4\mathring{\tau}L^{2})}{2^{2d-4}\,
    \mathring{\tau}L^{2d-3}}\bigg]\qquad 
\end{eqnarray}
with 
\begin{equation}
  \label{eq:Bd}
  B_d=\frac{\pi}{8\,\Gamma(3-d)\,\Gamma^2[(d-1)/2]\cos^{2}(d\pi/2)}\;.
\end{equation}

Aside from $I^{(\wp)}_{j}(L;\mathring{\tau})$ $(j=1,2)$, we also need
to calculate the integrals $J^{(\wp)}(L;\mathring{\tau})$ introduced
in Eq.~(\ref{eq:Jwp}) for $\wp=\mathrm{per}$ and $\wp=(\mathrm{sp},\mathrm{sp})$. To
this end we insert our above results for $I_{1}^{(\wp)}$ into
Eq.~(\ref{eq:Jwp}) and use the property (\ref{eq:diffeqQd2}),
obtaining
\begin{eqnarray}\label{eq:JperappzwR}
J^{\mathrm{(per)}}(L;\mathring{\tau}) & = &
\frac{2}{L}\frac{A_{d-1}}{d-1}\,\mathring{\tau}^{(d-1)/2}
-\frac{2A_{d}}{d}\,\mathring{\tau}^{d/2}\nonumber \\ 
 &  & \strut -\big(1-\lim_{\mathring{\tau}\to0}\big) \frac{8\pi
   Q_{d+2,2}(\mathring{\tau}L^{2})}{\mathring{\tau}L^{d+2}} \qquad
\end{eqnarray}
and
\begin{equation}\label{eq:relJperJspspapp}
J^{\mathrm{(sp,sp)}}(L;\mathring{\tau}) =
J^{\mathrm{(per)}}(2L;\mathring{\tau})\;.
\end{equation}

The $\mathring{\tau}\to0$ limit on the right-hand side can be evaluated in a
straightforward fashion with the aid of the 
representation
\begin{equation}
Q_{d,2}(r)=\frac{r^{(d+2)/4}}{(2\pi)^{d/2}}
\sum_{j=1}^{\infty}\frac{K_{(d-2)/2}\!\left(j\sqrt{r}\right)}{j^{(d-2)/2}},
\label{eq:Qd2sumBesselK}
\end{equation}
for $d\to d+2$, where  $K_\nu(x)$ is a modified Bessel function of the
second kind, and their well-known asymptotic behavior
\begin{equation}
K_\nu(x)\mathop{=}\limits_{x\to0}\;2^{\nu-1}\,
  \Gamma(\nu)\,x^{-\nu}+ O(x^{2-\nu})
\end{equation}
for $\nu>0$ (see, e.g., Eq.~(8.446) of Ref.~\cite{GR80}).

The result
\begin{equation}\label{eq:Qd2smally}
\lim_{r\to 0}\frac{8\pi\,
  Q_{d+2,2}(r)}{r}=
2\pi^{-d/2}\,\Gamma (d/2)\,\zeta(d)
\end{equation}
can be expressed in terms of either one of the one-loop Casimir
amplitude $\Delta^{(\wp)}_{C,[1]}$ given in Eq.~(\ref{eq:oneloopCasi}).
Inserting it into Eqs.~(\ref{eq:JperappzwR}) and
(\ref{eq:relJperJspspapp}) finally gives  
\begin{eqnarray}\label{eq:Jperres}
J^{\mathrm{(per)}}(L;\mathring{\tau}) & = &
-\frac{2\Delta^{\mathrm{(per)}}_{C,[1]}}{nL^{d}}
+\frac{2}{L}\frac{A_{d-1}}{d-1}\,\mathring{\tau}^{(d-1)/2}
\nonumber \\ 
 &  & \strut -\frac{2A_{d}}{d}\,\mathring{\tau}^{d/2}-\frac{8\pi
   Q_{d+2,2}(\mathring{\tau}L^{2})}{\mathring{\tau}L^{d+2}} \qquad
\end{eqnarray}
and
\begin{eqnarray}\label{eq:Jspspres}
J^{\mathrm{(sp,sp)}}(L;\mathring{\tau}) & = &
-\frac{2\Delta^{\mathrm{(sp,sp)}}_{C,[1]}}{nL^{d}}
+\frac{1}{L}\frac{A_{d-1}}{d-1}\,\mathring{\tau}^{(d-1)/2}
 \nonumber \\ 
 &  &\strut -\frac{2A_{d}}{d}\,\mathring{\tau}^{d/2}  -\frac{\pi\,
   Q_{d+2,2}(4\mathring{\tau}L^{2})}{2^{d-1}\,\mathring{\tau}L^{d+2}}\,,\qquad
\end{eqnarray}
respectively.

\section{Evaluation of $\bm{f_{\psi,[1]}^{(\wp)}(L;0)}$}\label{app:f10}

In this appendix we present the calculation of the one-loop free-energy
contributions $f_{\psi,[1]}^{(\wp)}(L;\mathring{\tau})$ defined in
Eq.~(\ref{eq:fpsi1}) for $\mathring{\tau}=0$. 
 
Upon applying Poisson's summation
formula~(\ref{eq:PoissonSumFormula}), we obtain
\begin{equation}
  \label{eq:fpsi1perapp}
  f_{\psi,[1]}^{(\mathrm{per})}(L;0)=f_{\psi,0}^{(\mathrm{per})}
  +nL\sum_{j=1}^{\infty}\int_{\bm{q}=(\bm{p},k)}^{(d)} \cos(kjL)\,\ln q^{2}
\end{equation}
and
\begin{equation}
  \label{eq:fpsi1spapp}
  f_{\psi,[1]}^{(\mathrm{sp},\mathrm{sp})}(L;0)=f_{\psi,0}^{(\mathrm{sp},\mathrm{sp})}
  +nL\sum_{j=1}^{\infty}
  \int_{\bm{q}=(\bm{p},k)}^{(d)}\cos(2kjL)\, \ln q^{2}\;,
\end{equation}
where
\begin{equation}
  \label{eq:fpsi0wpapp}
  f_{\psi,0}^{(\wp)}=L\,f_{b,0}+f_{\psi,s,0}^{(\wp)}
\end{equation}
with
\begin{equation}
  \label{eq:fbpsi0}
  f_{b,0}=\frac{n}{2} \int_{\bm{q}}^{(d)}\ln q^2
\end{equation}
and
\begin{equation}
  \label{eq:fswppsi0}
  f_{\psi,s,0}^{(\mathrm{per})}=2\,f_{\psi,s,0}^{(\mathrm{sp},\mathrm{sp})}
  =-\frac{n}{2}\, \int_{\bm{p}}^{(d-1)}\ln p^{2}\;.
\end{equation}

The expressions~(\ref{eq:fpsi1perapp}) and
(\ref{eq:fpsi1spapp}) can be evaluated along lines similar to those
followed in Sec.~V. of Ref.~\cite{KD92a}. This gives
\begin{equation}
  nL\sum_{j=1}^{\infty} \int_{\bm{q}=(\bm{p},k)}^{(d)}\cos(kjL)\ln q^{2}
  =L^{-(d-1)}\Delta_{C,[1]}^{(\mathrm{per})},
\end{equation}
and
\begin{equation}
 n L\sum_{j=1}^{\infty}\int_{\bm{q}=(\bm{p},k)}^{(d)}  \cos(2kjL)\ln q^{2}
  =L^{-(d-1)}\Delta_{C,[1]}^{(\mathrm{sp},\mathrm{sp})}, 
\end{equation}
where $\Delta_{C,[1]}^{(\wp)}$ are the one-loop Casimir amplitudes
of Eq.~(\ref{eq:oneloopCasi}).

\section{Series representations of the functions $\bm{Q_{d,\sigma}(r)}$}
\label{app:Qdsigma} 

In this appendix we wish to derive representations of the functions
$Q_{d,\sigma}(r)$ as generalized power series and to establish their
relation~(\ref{eq:Qd2ga}) with the functions $g_{a,b}(z)$ utilized by KD.

Let us define the integral
\begin{equation}
I_{d,\sigma}(k,y)=K_{d-1}\int_{0}^{\infty} dp\,
p^{d-2}\,\frac{(k^{2}
  +p^2)^{(\sigma-2)/2}}{y+k^{2}+p^2},
\end{equation}
where $K_d$ denotes the usual factor
\begin{equation}
K_d\equiv\int_{\bm{q}}^{(d)}\,\delta(|\bm{q}|-1)=
\frac{2^{1-d}\,\pi^{-d/2}}{\Gamma(d/2)}.
\end{equation}
Then the right-hand side of
Eq.~(\ref{eq:definitionQdsigma}) can be written as
\begin{eqnarray}\label{eq:decompQdsigma}
Q_{d,\sigma}(r) & = & \frac{r}{2}\bigg\{ \big[I_{d,\sigma}(0,1)
    -I_{d+1,\sigma}(0,1)\sqrt{r}\big]\,r^{(d+\sigma-5)/2}\nonumber \\
 &  & \qquad\strut +\sum_{k\ne 0}I_{d,\sigma}(k,r)\bigg\} ,
\end{eqnarray}
where here and below the summation $\sum_{k\ne 0}$  extends over all
nonzero $k\in2\pi\mathbb{Z}$.

When $k=0$, the evaluation of the integral $I_{d,\sigma}(k,r)$ is
straightforward, giving
\begin{equation}
I_{d,\sigma}(0,r)=\frac{2^{1-d}\,\pi^{(3-d)/2}\,r^{(d+\sigma-5)/2}}{\Gamma[
  (d-1)/2]
  \cos[\pi(d+\sigma)/2]}. 
\label{eq:Idsigma}
\end{equation}

In the case of nonzero values of $k$, we Taylor expand in
$r$ about $r=0$ to obtain
\begin{equation}
I_{d,\sigma}(k,r)=\sum_{j=0}^{\infty}I_{d,\sigma}^{(0,j)}(k,0)\,
\frac{r^j}{j!} \;, 
\label{eq:seriesIdsigma}
\end{equation}
where the required partial derivatives
$I_{d,\sigma}^{(0,j)}(k,r)=\partial^jI_{d,\sigma}(k,r)/\partial r^j$
at $r=0$ are given by
\begin{equation}
I_{d,\sigma}^{(0,j)}(k,0)=\frac{(-1)^j
  \Gamma(j+1)\,
  \Gamma[j+(5-d-\sigma)/2]}{(4\pi)^{(d-1)/2}\,|k|^{2j+5-d-\sigma}\, 
  \Gamma(j+2-\sigma/2)}. 
\label{eq:Inullnderivatives}
\end{equation}

We now substitute the Taylor series~(\ref{eq:seriesIdsigma}) into
Eq.~(\ref{eq:decompQdsigma}) and interchange the summations over $j$
and $k$ in the last term. Recalling the series expansion
\begin{equation}
  \zeta(s)=\sum_{n=1}^{\infty}\frac{1}{n^{s}},
\end{equation}
of Riemann's zeta function, one can perform the $k$-summations
analytically to obtain
\begin{equation}
  \sum_{k\neq 0}I_{d,\sigma}^{(0,j)}(k,0)=
  2\,I_{d,\sigma}^{(0,j)}(2\pi,0)\,\zeta(2j+5-d-\sigma).
\label{eq:sumIdsigmaderiv2}
\end{equation}
Using this result together with Eq.~(\ref{eq:Idsigma}), one arrives at
the representation
\begin{eqnarray}\label{eq:Qdsiggps}
  Q_{d,\sigma}(r)
  & = & \frac{2^{-d}\,\pi^{(3-d)/2}}{\Gamma[(d-1)/2]\,\cos[\pi(d+\sigma)/2]}\,r^{(d+\sigma-3)/2} \nonumber\\ &&
\strut +\frac{2^{-d-1}\,\pi^{1-d/2}}{\Gamma(d/2)
    \sin[\pi(d+\sigma)/2]} \,r^{(d+\sigma-2)/2}\nonumber \\
  & &
  \strut -\pi^{(d-1)/2}\sum_{j=1}^{\infty}\bigg[ \frac{\Gamma[j+(3-d-\sigma)/2
    ]}{
     \Gamma(j+1-\sigma/2)}\nonumber\\
&&\qquad\times
\frac{\zeta(2j+3-d-\sigma)}{(2\pi)^{2j-\sigma+2}}\,(-r)^{j}\bigg].
\label{eq:genpowserQdsigma}
\end{eqnarray}

For general values of $d$ and $\sigma$, the first two terms on the
right-hand side of Eq.~(\ref{eq:Qdsiggps}) have branch-cut
singularities. Cauchy's ratio test shows that the remaining power
series (3rd term) is absolutely convergent for complex $y$ inside a
circle of radius $(2\pi)^{2}$.

As is known from Ref.~\cite{DDG06}, functions $Q_{d,\sigma}$ with
noninteger values of $\sigma$ are encountered in the study of
finite-size effects of systems with long-range interactions. From the
series expansion~(\ref{eq:Qdsiggps}) the asymptotic behavior of the
functions $Q_{d,\sigma}(r)$ as $r\to 0$ can be read off easily even
for such general values of $\sigma$.  This representation may, of
course, also be employed to compute the functions $Q_{d,\sigma}(r)$ by
numerical means for values of $r$ inside the radius of convergence of
the series.

To establish the relation~(\ref{eq:Qd2ga}) between $Q_{d,\sigma}$ and
the functions $g_{a,0}(z)$ [cf.  Eq.~(\ref{eq:ga})] employed by KD, it
is convenient to use the expansion~(\ref{eq:Qd2sumBesselK}).
Substituting the Bessel functions in it by their integral
representation
\begin{equation}
  \label{eq:Knuintrep}
  K_\nu(z)=\frac{(z/2)^\nu\,\sqrt{\pi}}{\Gamma(\nu+\frac{1}{2})}\,
  \int_1^\infty dt\, (t^2-1)^{\nu-\frac{1}{2}}\,e^{-z\,t}
\end{equation}
and interchanging the integration with the summation over $j$
immediately gives Eq.~(\ref{eq:Qd2ga}).

\section{Numerical results for and properties of the required functions $\bm{Q_{d,\sigma}(r)}$}
\label{app:Qd2}

In the present work only functions $Q_{d,\sigma}(r)$ with the special
value $\sigma=2$ are needed. The purpose of the present appendix is to
present numerical results for these functions.

\begin{figure}[htb]
  \centering
\includegraphics[width=0.85\columnwidth,clip]{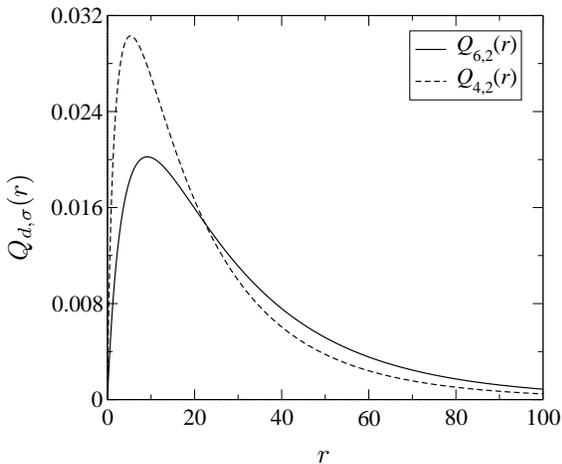}
  \caption{Plots of the functions $Q_{4,2}(r)$ and $Q_{6,2}(r)$,
obtained by numerical evaluation of the series
expansion~(\ref{eq:Qd2sumBesselK}).
\label{fig:Qd2}}
\end{figure}

The expansion~(\ref{eq:Qd2sumBesselK}) of these functions in terms of
modified Bessel functions lends itself well to numerical evaluation.
Figure~\ref{fig:Qd2} shows plots of the functions 
$Q_{4,2}(r)$ and $Q_{6,2}(r)$, which were numerically determined via
this representation.

\begin{figure}[htb]
\begin{center}\includegraphics[width=0.85\columnwidth,clip]{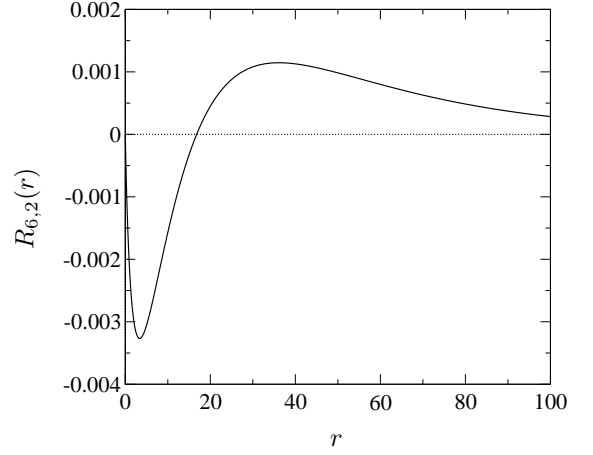}
\end{center}
\caption{Plot of the function $R_{6,2}(r)$ as numerically obtained
  from the series expansion that results from Eq.~(\ref{eq:Qd2sumBesselK}) 
  by differentiation with respect to $d$. \label{fig:R62}}
\end{figure}

The function $R_{d,2}(r)$ has an expansion in modified Bessel
functions analogous to Eq.~(\ref{eq:Qd2sumBesselK}), which follows
from it by differentiation with respect to $d$. Using it we have
determined $R_{6,2}(r)$ by numerical evaluation. The result is
depicted in Fig.~\ref{fig:R62}.

The asymptotic small-$r$ forms of these functions can be determined in
a straightforward fashion from the representation
(\ref{eq:Qdsiggps}). One obtains
\begin{eqnarray}\label{Q42smallr}
  \frac{Q_{4,2}(r)}{r}&\mathop{=}\limits_{r\to 0}&
\frac{1}{24}-\frac{\sqrt{r}}{8 \pi } -\frac{r\ln r}{32 \pi ^2}
\nonumber\\ &&\strut 
+\frac{1-2\,
   C_E +2\ln(4 \pi)}{32 \pi
   ^2}\,r
+\frac{\zeta (3) }{256 \pi ^4}\,r^2
\nonumber\\ &&\strut
-\frac{\zeta (5) }{2048 \pi ^6}\,r^3+O(r^4)\;,
\end{eqnarray}
\begin{eqnarray}\label{Q62smallr}
  \frac{Q_{6,2}(r)}{r}&\mathop{=}\limits_{r\to 0}&
\frac{\pi }{360}-\frac{r}{96 \pi }
+\frac{r^{3/2}}{48 \pi ^2}+\frac{r^2\ln r}{256\pi^3}\nonumber\\
&&\strut -\frac{3/2-2C_E+2\ln(4\pi)}{256\pi^3}\,r^2+O(r^3)\;,\qquad
\end{eqnarray}
and
\begin{eqnarray}\label{R622smallr}
  \frac{R_{6,2}(r)}{r}&\mathop{=}\limits_{r\to 0}& \frac{ C_E + \ln(4\pi
    )-240\, \zeta '(-3)-8/3}{720}\,\pi \nonumber\\ &&
  \strut-\frac{C_E -2+24 \,\zeta'(-1) +\ln (4
    \pi )}{192 \pi }\,r \nonumber\\ && \strut +\frac{\ln (r/\pi)+
    C_E -8/3 }{96 \pi
   ^2}\,r^{3/2}
\nonumber\\ && \strut
+O\left(r^2\ln^2r\right).
\end{eqnarray}

Their asymptotic forms for large values of $r$ follow from (cf.\ equation
(B33) of Ref.~\cite{DDG06})
\begin{equation}
  \label{eq:Qd2largeas}
  Q_{d,2}(r)\mathop{=}\limits_{r\to
    \infty}\frac{r^{(d+1)/4}}{2\,(2\pi)^{(d-1)/2}}\, e^{-\sqrt{r}} \,\Big[1+
  O\big(r^{-1/2}\big)\Big]\;,
\end{equation}
giving
\begin{equation}\label{Q42larger}
   \frac{Q_{4,2}(r)}{r}\mathop{=}\limits_{r\to \infty}
\frac{r^{1/4}}{2\,
  (2\pi)^{3/2}}\,e^{-\sqrt{r}}\,\Big[1+O\big(r^{-1/2}\big)\Big]\;, 
\end{equation}
\begin{equation}\label{Q62larger}
   \frac{Q_{6,2}(r)}{r}\mathop{=}\limits_{r\to \infty}
\frac{r^{3/4}}{2\, (2\pi)^{5/2}}\,e^{-\sqrt{r}}\, \Big[1
+O\big(r^{-1/2} \big)\Big]\;,
\end{equation}
and
\begin{equation}\label{R62larger}
  \frac{R_{6,2}(r)}{r}\mathop{=}\limits_{r\to\infty} \frac{r^{3/4}}{8\,(2\pi)^{5/2}} \,e^{-\sqrt{r}} \,\ln\frac{r}{4 \pi ^2}\Big[1
+O\big(r^{-1/2} \big)\Big]   \;,
\end{equation}
respectively.


\begin{thebibliography}{61}
\bibitem[{rem({\natexlab{c}})}]{rem:corrlength}
\bibinfo{note}{Owing to the difference between our choice of $N_d$ and the
  alternative one employed in the review article \cite{Die86a} and by KD, the
  non-universal amplitude $\xi_\infty^+=\lim_{\tau\to
  0+}\xi_\infty|_{u=u^*}\,\tau^\nu$ of the second-moment bulk correlation
  length $\xi_\infty$ is unity at first order in $\epsilon$, i.e.,
  $\xi_\infty^+=1+O(\epsilon^2)$. The latter is defined in the conventional
  manner, see Eq.~(\ref{eq:xiinfdef}).}
\bibitem[{rem({\natexlab{d}})}]{rem:nonanalytcsp}
\bibinfo{note}{For fixed $d=4-\epsilon$, the special value
  $\mathring{c}_{\text{sp}}\sim \mathring{u}^{1/\epsilon}$ depends in a
  non-analytic manner on the coupling constant $\mathring{u}$. Just as the bulk
  critical value $\mathring{\tau}_{c,\infty}$, it cannot therefore be
  determined by perturbation theory; for details, see Refs.~\cite{DS94,DS98}.}
\bibitem[{rem({\natexlab{e}})}]{rem:KD}
\bibinfo{note}{Our choice of additive counterterms differs from that of KD, who
  chose them $L$-dependent, fixing them through normalization conditions for
  the $L$-dependent free energy. A second difference is that KD restricted
  themselves to the fixed-point values $c_j=0,\infty$ and hence did not take
  into account $c$-dependent additive surface counterterms.}
\bibitem[{rem({\natexlab{f}})}]{rem:sfdd}
\bibinfo{note}{Note that subtracting $T^{\le
  4}_{\text{NP}}f_b(\mathring{\tau},\mathring{u})$ from $f_b$ and $T^{\le
  3}_{\text{NP}} f_s(\mathring{\tau}, \mathring{u},\mathring{c})$ from $f_s$
  reduces their superficial degrees of divergence by $4$ and $3$, respectively,
  so that the differences become superficially convergent when $d\le 4$.}
\bibitem[{rem({\natexlab{g}})}]{rem:Ec}
\bibinfo{note}{Since the flow of $c_j$ is the same as for the corresponding
  semi-infinite systems, the same nonuniversal metric factor, denoted
  $E_c^*(u)$ in Ref.~\cite{Die86a}, is associated with either one of $c_1$ and
  $c_2$.}
\bibitem[{rem({\natexlab{h}})}]{rem:RGvert}
\bibinfo{note}{This follows from the fact that correlation functions of the
  $\psi$ field as well as the field $\varphi$ are multiplicatively
  renormalizable.}
\bibitem[{rem({\natexlab{i}})}]{rem:surfaniso}
\bibinfo{note}{In the presence of easy-axis spin anisotropies at the surface a
  multicritical point at which a line of $d=2$ surface transitions of the Ising
  type reaches the line of $n\ge 2$ bulk transitions. The associated
  anisotropic special transitions were investigated in Ref.~\cite{DE82,DE84}.
  Such surface anisotropies will not be considered here.}
\end{thebibliography}

\end{document}